\documentclass[aps,preprintnumbers,amsmath,nofootinbib]{revtex4}
\usepackage{amssymb,amsthm}
\usepackage{graphicx}
\usepackage{dcolumn}
\usepackage{footnpag}
\usepackage{color}

\begin{document}

%%%%%%%%%%%%%%%%%%%%%%%%%%%%%%%%%%%%%%%%%%%%%%%%%%%%%%%%%%%%%%%%%%%%%%%%%%%%%%
\title{L\'{e}vy distributions for one-dimensional analysis of the
Bose\,--\,Einstein correlations}
%%%%%%%%%%%%%%%%%%%%%%%%%%%%%%%%%%%%%%%%%%%%%%%%%%%%%%%%%%%%%%%%%%%%%%%%%%%%%%

\author{V.A. Okorokov} \email{VAOkorokov@mephi.ru; Okorokov@bnl.gov}
\affiliation{National Research Nuclear University MEPhI (Moscow
Engineering Physics Institute), Kashirskoe shosse 31, 115409 Moscow, Russian Federation}

\date{\today}

\begin{abstract}
A general study of relations between the parameters of two centrally-symmetric L\'{e}vy distributions, often used for one-dimensional investigation of Bose\,--\,Einstein correlations, is given for the first time. These relations of the strength of correlations and of the radius of the emission region take into account possible various finite ranges of the Lorentz invariant four-momentum difference for two centrally-symmetric L\'{e}vy distributions. In particular, special cases of the relations are investigated for Cauchy and normal (Gaussian) distributions. The mathematical formalism is verified using the recent measurements given a generalized centrally-symmetric L\'{e}vy distribution is used. The reasonable agreement is observed between estimations and experimental results for all available types of strong interaction processes and collision
energies.
\vspace*{0.5cm}

% end of PACS codes
\end{abstract}

\maketitle
%#1
\section{Introduction}\label{sec:1}
Correlations between two identical bosons, called
Bose\,--\,Einstein correlations (BEC), are a well-known phenomenon in
high-energy and nuclear physics. These correlations play an
important role in the studies of multiparticle production and soft
physics. Constructive interference affects the joint probability
for the emission of a pair of identical bosons with four-momenta
$p_{1}$ and $p_{2}$. Experimentally, the one-dimensional BEC
effect is observed as an enhancement at low values of the Lorentz
invariant quantity $q=\sqrt{-(p_{1}-p_{2})^{2}} \geq 0$ in the
two-particle correlation function (CF),
\begin{equation*}
\mathbf{C}_{2}(q) = \rho(q) / \rho_{\scriptsize{\mbox{ref}}}(q).
\end{equation*}
Here the $\rho$ is the two-particle density function,
$\rho_{\scriptsize{\mbox{ref}}}$ is a reference two-particle
density function that by construction is expected to include no
BEC. The detailed shape analysis of the peak of CF is an important
topic on theoretical and experimental points of view because this
shape carries information about the possible features of
space-time structure of particle source
\cite{Kittel-APPB-32-3927-2001,Csorgo-HIP-15-1-2002}. For
instance, the detail investigations have to do for shape of
correlation peak in modern experiments with high statistics for
verification of hypothesis of possible self-affine fractal-like
geometry of emission region \cite{Okorokov-arXiv-1312.4269,Okorokov-AHEP-2016-5972709-2016}.
The BEC effect in one dimension is usually described by a
few-parameter function for which several different functional
forms have been proposed.
The power-law parametrization $\mathbf{C}_{2}(q) \sim q^{-\beta}$ is the important signature for fractal-like source extending over a large volume \cite{Bialas-APPB-23-561-1992,Bialas-NPA-545-285c-1992}. The quite reasonable fit is achieved with this parametrization of two-pion CF in various types of multiparicle production processes \cite{Kittel-APPB-32-3927-2001}. But unfortunately power-law fits are absent for high-statistics modern experimental data so far \cite{Okorokov-PowerFit-inprogress}. On the other hand the stable (on L\'{e}vy) distributions \cite{Levy-1937} are one of the most
promising tools for studies of fractal-like space-time extent of emission region. These distributions are a rich
class of probability distributions that allow skewness and heavy
tails and have many important physical applications. As shown in
\cite{Okorokov-arXiv-1312.4269,Okorokov-AHEP-2016-5972709-2016} the subclass of non-isotropic
centrally-symmetric L\'{e}vy distributions
\cite{Uchaikin-ZETF-124-903-2003,Uchaikin-UFN-173-847-2003} is most useful for studies of
Bose\,--\,Einstein CF. Therefore this subclass of centrally-symmetric L\'{e}vy distributions is considered regarding of BEC measurements in the present paper .

For low-dimensional (1D) analysis the
centrally-symmetric L\'{e}vy distribution results in the most
general parametrization of the experimental Bose\,--Einstein CF
\begin{equation}
\mathbf{C}_{2}(q) \propto 1 +
\Omega(\alpha,\lambda,z),~~~~~\Omega(\alpha,\lambda,z) \equiv
\lambda \exp(-|z|^{\alpha}),~~~~~z \equiv qR. \label{eq:1.1}
\end{equation}
Here $\lambda$ is the strength of correlations called also
chaoticity, $R$ is the 1D BEC radius, $0 < \alpha \leq 2$ is the
L\'{e}vy index called also index of stability. As
known for a static source with no
final state interactions \cite{Goldhaber-PR-120-300-1960,Boal-RMP-62-553-1990}, there is the relation
$\mathbf{C}_{2}(q) \propto |\tilde{f}(q)|^{2}$, $\tilde{f}(q)=\displaystyle\int\!dx\exp(iqx)f(x)$, i.e. Bose\,--\,Einstein CF $\mathbf{C}_{2}(q)$ measures the absolute value squared Fourier
transformed source density in the coordinate space $f(x)$, $\displaystyle\int\!dxf(x)=1$ called also
coordinate-space distribution function of the particle emission
points. The various experiments use the different forms of
the (\ref{eq:1.1}) which correspond to the various hypotheses with
regard of $f(x)$. For example, most of the earliest experiments
with particle beams used the specific case of the (\ref{eq:1.1})
at $\alpha=2$ -- the Gaussian parametrization corresponded to the
normal (Gaussian) distribution function
$f_{\footnotesize{G}}(x)=(2\pi
R^{2}_{\footnotesize{G}})^{-1/2}\exp\bigl[-(x-x_{0})^{2}/2R^{2}_{\footnotesize{G}}\bigr]$,
where the Gaussian scale parameter is
$R^{2}_{\footnotesize{G}}=\langle x^{2}\rangle-x_{0}^{2}$, the
standard deviation; then another specific case of the
(\ref{eq:1.1}) at $\alpha=1$ is used widely, especially, for
particle (not nuclear) collisions. The equation (\ref{eq:1.1}) at
$\alpha=1$ is called exponential parametrization for Bose\,--\,Einstein CF
$\mathbf{C}_{2}(q)$ and it corresponds to the Cauchy (Lorentzian)
distribution function
$f_{\footnotesize{C}}(x)=\pi^{-1}R_{\footnotesize{C}}/[R_{\footnotesize{C}}^{2}+(x-x_{0})^{2}]$
with scale parameter $R_{\footnotesize{C}}$
\cite{Csorgo-EPJC-36-67-2004,Forbes-book-2010}.
Futhermore the recent studies at the LHC
\cite{CMS-PRL-105-032001-2010,CMS-JHEP-0511-029-2011,Padula-WPCF-2014,ATLAS-Astalos-PhD2015}
demonstrate that general view of the (\ref{eq:1.1}) allows the
reasonable description of experimental CF, particular for
proton-proton ($p+p$) collisions but for centrally-symmetric L\'{e}vy distribution with $\alpha \in (0;2)$, $\alpha \ne 1$ the corresponding source density in coordinate space $f(x)$ can be written analytically for
$\alpha=3/2, 2/3, 1/2, 1/3$ only \cite{Uchaikin-UFN-173-847-2003}. It is often difficult to compare results from different experiments because of the many different data analysis methods \cite{Boal-RMP-62-553-1990}, in particular due to various parameterizations for 1D Bose\,--\,Einstein CF $\mathbf{C}_{2}(q)$. Therefore the derivation of the
relations between the sets of BEC parameters for two
centrally-symmetric L\'{e}vy distributions is the important task
for correct comparison of the results from different experiments,
creation of the global kinematic (energy, pair transverse momentum
etc.) dependencies of BEC parameters and so on. Such studies are important for investigations of common features of soft-stage dynamics in various
multiparticle production processes as well as for equation of
state (EoS) of strongly interacting matter, in particular, search
for phase transition to the quark-gluon deconfined matter.
It would be noted the study of energy dependence of pion BEC parameters in heavy ion collisions \cite{Adler-PRL-87-082301-2001} was one of the main causes and drivers for hypothesis of cross-over transition from strongly-coupled quark-gluon phase to hadronic one at Relativistic Heavy Ion Collider (RHIC) energies $\sqrt{s_{\footnotesize{NN}}} \sim 100$ GeV. Furthermore some results for deconfinement in small system
\cite{Abelev-PRC-79-034909-2009,Gutay-IJMPE-24-1550101-2015} indicate remarkable similarity
of both the bulk and the thermodynamic properties of strongly interacting matter created
in high energy $p+p$ / $\bar{p}+p$ and $\mbox{A}+\mbox{A}$ collisions. The BEC can provide new knowledge about collectivity and possible creation of droplets of quark-gluon matter in small system collisions. For these studies the correct comparison can be crucially important for BEC parameters in various multiparticle processes for wide energy range. But as mentioned above Bose\,--\,Einstein CF $\mathbf{C}_{2}(q)$ is often described by different view of the (\ref{eq:1.1}) depending on type of reaction, collision energy and features of experimental analysis. Therefore the study of centrally-symmetric L\'{e}vy distributions and search for relations between parameters for corresponding CF has scientific interest for physics of strong interactions.

The paper is organized as follows. In Sec.\,\ref{sec:2},
mathematical formalism is described for case of two general view
centrally-symmetric L\'{e}vy distributions. Dependencies of
desired 1D BEC observables on $q$ and $\alpha$ are studied for
\emph{a priori} known parameters for second centrally-symmetric
L\'{e}vy distributions. Section\,\ref{sec:3} is devoted to the
detail discussion of specific case of these distributions, namely,
Cauchy and Gaussian ones most used in experimental investigations
of 1D CF $\mathbf{C}_{2}(q)$. Database of experimental results for
set of 1D BEC parameters $\{\lambda, R\}$ for charged pion source
in strong interaction processes is created within the framework of
the paper in order to verify the mathematical formalism.
Sec.\,\ref{sec:4} demonstrates the comparison between the
estimations calculated for 1D BEC parameters with help of
mathematical formalism under discussion and available experimental
results for various reactions and in wide energy range\footnote{In
should be noted that in Sec.\,\ref{sec:3} and \ref{sec:4} the 1D
BEC parameters are supplied with the subindexes according with the
names of corresponding source distribution function, namely, "L"
is for the general view of centrally-symmetric L\'{e}vy
distribution, "C" is for the Cauchy source distribution function
and "G" -- for Gaussian one. Otherwise the notations
$\{\lambda_{\footnotesize{E}}, R_{\footnotesize{E}}\}$ are often
used in papers for second case due to relation between Cauchy
distribution for $f(x)$ and exponential parametrization for
Bose\,--\,Einstein CF $\mathbf{C}_{2}(q)$ discussed above. As
consequence the mathematically rigorous terminology is used over
full manuscript: the term "Cauchy distribution" corresponds to the
source function in coordinate space $f_{\footnotesize{C}}(x)$ and
the term "exponential function / parametrization" is used for the
related parametrization of correlation function
$\mathbf{C}_{2,\footnotesize{E}}(q)$; for the case of arbitrary $0
< \alpha < 2$, $\alpha \ne 1$ the term "centrally-symmetric
L\'{e}vy" is suitable for both the source function in coordinate
space $f_{\footnotesize{L}}(x)$ and the parametrization of
correlation function $\mathbf{C}_{2,\footnotesize{L}}(q)$; the
similar situation is for $\alpha=2$: the term "Gaussian" is
applicable for both the source function in coordinate space
$f_{\footnotesize{G}}(x)$ and the corresponding parametrization of
correlation function $\mathbf{C}_{2,\footnotesize{G}}(q)$.}. In
Sec.\,\ref{sec:5} some final remarks are presented. The
experimental database is shown in the Appendix\,\ref{sec:6} for 1D
BEC parameters.

%#2
\section{Relations between BEC parameters in general case}\label{sec:2}

Let some experimental CF $\mathbf{C}_{2}(q)$ is described by two
parameterizations (\ref{eq:1.1}) with $\Omega_{1} \equiv
\Omega(\alpha_{1},\lambda_{1},z_{1})$ and $\Omega_{2} \equiv
\Omega(\alpha_{2},\lambda_{2},z_{2})$. Then relations between
parameters of $\Omega_{1}$ and $\Omega_{2}$ can be deduced on the
basis that both parameterizations describe one experimental CF
$\mathbf{C}_{2}(q)$, i.e. one sample of experimental
points\footnote{In general the approximations of
$\mathbf{C}_{2}(q)$ are characterized by different qualities for
various parameterizations (\ref{eq:1.1}) with $\Omega_{1}$ and
$\Omega_{2}$. The influence of this difference is not studied in
present work and can be considered as separate task.}. Thus one
can assume that the areas under fit curves for two
parameterizations (\ref{eq:1.1}) with $\Omega_{1}$ and
$\Omega_{2}$ are approximately equal to each other as well as the
first moments of the corresponding centrally-symmetric L\'{e}vy
distributions.

\subsection{Mathematical formalism}\label{subsec:2.1}
The relations between two sets of parameters $\{\lambda_{1},R_{1}\}$ and
$\{\lambda_{2},R_{2}\}$ of the particle source can be derived from
the following system of equations:
\begin{subequations}
\begin{equation}
S_{1} = S_{2},~~~\forall~i=1,2: S_{i} \equiv \int_{J_{i}}
\Omega(\alpha_{i},\lambda_{i},z_{i})dq; \label{eq:2.1.a}
\end{equation}
\vspace*{-0.5cm}
\begin{equation}
\langle q\rangle_{1} = \langle q\rangle_{2},~~~\forall~i=1,2:
\langle q\rangle_{i} \equiv S_{i}^{-1} \int_{J_{i}}
q\Omega(\alpha_{i},\lambda_{i},z_{i})dq. \label{eq:2.1.b}
\end{equation} \label{eq:2.1}
\end{subequations}
\hspace*{-0.12cm}The first equation (\ref{eq:2.1.a}) corresponds
to the equality of the areas under fit curves and the
(\ref{eq:2.1.b}) is the equality of the first moments of the
$\Omega(\alpha_{i},\lambda_{i},z_{i})$ distributions\footnote{As
discussed above the approximate equalities are expected for areas
and first moments in general case. This softer condition is enough
for applicability of the formalism suggested in the paper. But the
exactly equal signs are used in the (\ref{eq:2.1}) as well as in
the text below in order to get the mathematically correct forms
for the systems of equations.}. The system (\ref{eq:2.1}) contains
the equations allow the estimation of unknown strength of
correlations and 1D BEC radius based on the available values of
these parameters but it supposes the L\'{e}vy indexes $\alpha_{i}$
are known \emph{a priori} for both parameterizations $\Omega_{i}$,
$i=1,2$. In equations (\ref{eq:2.1}) the integrals are taken over
full fit ranges $J_{i}$, $i=1,2$ for corresponding
parameterizations with $\Omega_{i}$ of experimental CF. It should
be noted that in general case (i) the ranges of integration can be
different for parameterizations with $\Omega_{1}$ and
$\Omega_{2}$; (ii) the full fit range can be the set of subranges
due to possible exception of some intervals of the relative
4-momentum (regions of resonance contributions etc.), i.e.
$\forall\,i=1,2:
J_{i}=\displaystyle{\bigcup}_{k=1}^{N_{i}}J_{i}^{k} \equiv
\displaystyle{\bigcup}_{k=1}^{N_{i}}[q_{k,\scriptsize{\mbox{min}}},q_{k,\scriptsize{\mbox{max}}}]$
and consequently for all types of integrals and $\forall\,i=1,2$
in the system (\ref{eq:2.1}): $\displaystyle{\int_{J_{i}}} \to
\displaystyle{\sum}_{k=1}^{N_{i}}\displaystyle{\int_{J_{i}^{k}}}$.
But usually the fit ranges $J_{i}$ are identical for both
$\Omega_{i}$, $i=1,2$ in experimental studies (see, for example,
\cite{CMS-PRL-105-032001-2010}). In general case of the
centrally-symmetric L\'{e}vy distributions and finite fit ranges
the system equations under consideration can not be solved
analytically. The numerical procedure should be used in order to
get the relations between two sets of parameters
$\{\lambda_{1},R_{1}\}$ and $\{\lambda_{2},R_{2}\}$ of the
particle source in this case. Without loss of generality the
$\Omega_{1}$ and $\alpha_{2}$ are considered as known and values
of BEC parameters $\{\lambda_{2},R_{2}\}$ are supposed as desired
below. Then for specific case of semi-infinite ranges for
integration $\forall\,i=1,2: J_{i}=[0;\infty)$ the system
(\ref{eq:2.1}) can be solved analytically and one can derive the
following ultimate relations between two sets
$\{\lambda_{1},R_{1}\}$ and $\{\lambda_{2},R_{2}\}$ of BEC
parameters for corresponding centrally-symmetric L\'{e}vy
parameterizations with $\Omega_{i}$, $i=1,2$
\begin{subequations}
\begin{equation}
\lambda_{1}^{u} =
\lambda_{2}\bigl[\alpha_{1}\Gamma(2\alpha_{1}^{-1})\Gamma^{2}(\alpha_{2}^{-1})\bigr]
\bigl[\alpha_{2}\Gamma^{2}(\alpha_{1}^{-1})\Gamma(2\alpha_{2}^{-1})\bigr]^{-1};
\label{eq:2.2.a}
\end{equation}
\vspace*{-0.5cm}
\begin{equation}
R_{1}^{u} =
R_{2}\bigl[\Gamma(2\alpha_{1}^{-1})\Gamma(\alpha_{2}^{-1})\bigr]
\bigl[\Gamma(\alpha_{1}^{-1})\Gamma(2\alpha_{2}^{-1})\bigr]^{-1}\label{eq:2.2.b}
\end{equation}\label{eq:2.2}
\end{subequations}
\hspace*{-0.12cm}and vice versa. Here $\displaystyle
\Gamma(x)=\int^{\infty}_{0}t^{x-1}\exp(-t)dt,~\mbox{Re}\,x
> 0$ is the gamma function.

In the point of view of data analysis the absence of the general
analytic relations between $\{\lambda_{1},R_{1}\}$ and
$\{\lambda_{2},R_{2}\}$ leads to the following approach for
estimations of the errors of the unknown parameters. Let without
the loss of generality suppose that the values are known for set
of parameters $\{\alpha_{1},\lambda_{1},R_{1}\}$ with its errors
$\{\Delta^{\pm}\alpha_{1},\Delta^{\pm}\lambda_{1},\Delta^{\pm}R_{1}\}$
for centrally-symmetric L\'{e}vy parametrization with $\Omega_{1}$
as well as for $\alpha_{2}$ with $\Delta^{\pm}\alpha_{2}$ for
parametrization with $\Omega_{2}$. The two sets of values for
unknown BEC parameters $\{\lambda_{2},R_{2}\}$ can be obtained
with the help of suitable system of equations: the input values
$\{\alpha_{1}+\Delta^{+} \alpha_{1},\lambda_{1}+\Delta^{+}
\lambda_{1},R_{1}+\Delta^{+}R_{1}\}$ and $\alpha_{2}+\Delta^{+}
\alpha_{2}$ produce the output set $\{\lambda_{2}^{+},R_{2}^{+}\}$
and $\{\alpha_{1}-\Delta^{-} \alpha_{1}, \lambda_{1}-\Delta^{-}
\lambda_{1},R_{1}-\Delta^{-}R_{1}, \alpha_{2}-\Delta^{-}
\alpha_{2}\} \to \{\lambda_{2}^{-},R_{2}^{-}\}$. Then the error
estimations for set $\{\lambda_{2},R_{2}\}$ of BEC parameters for
parametrization with $\Omega_{2}$ can be calculated as follows:
\begin{equation}
\Delta^{\pm} Y_{2} = |Y_{2}^{\pm} - Y_{2}|, Y_{2} \equiv
\lambda_{2}, R_{2}. \label{eq:2.3}
\end{equation}
One can use the errors (\ref{eq:2.3}) which are asymmetric in
general case or make the averaging of up and low uncertainties and
then to use the symmetric errors $\Delta Y_{2} = (\Delta^{+} Y_{2}
+ \Delta^{-} Y_{2})/2$.

\subsection{Dependencies on $q$ and $\alpha$
variables}\label{subsec:2.2} 
\begin{figure*}
\includegraphics[width=15.5cm,height=17.0cm]{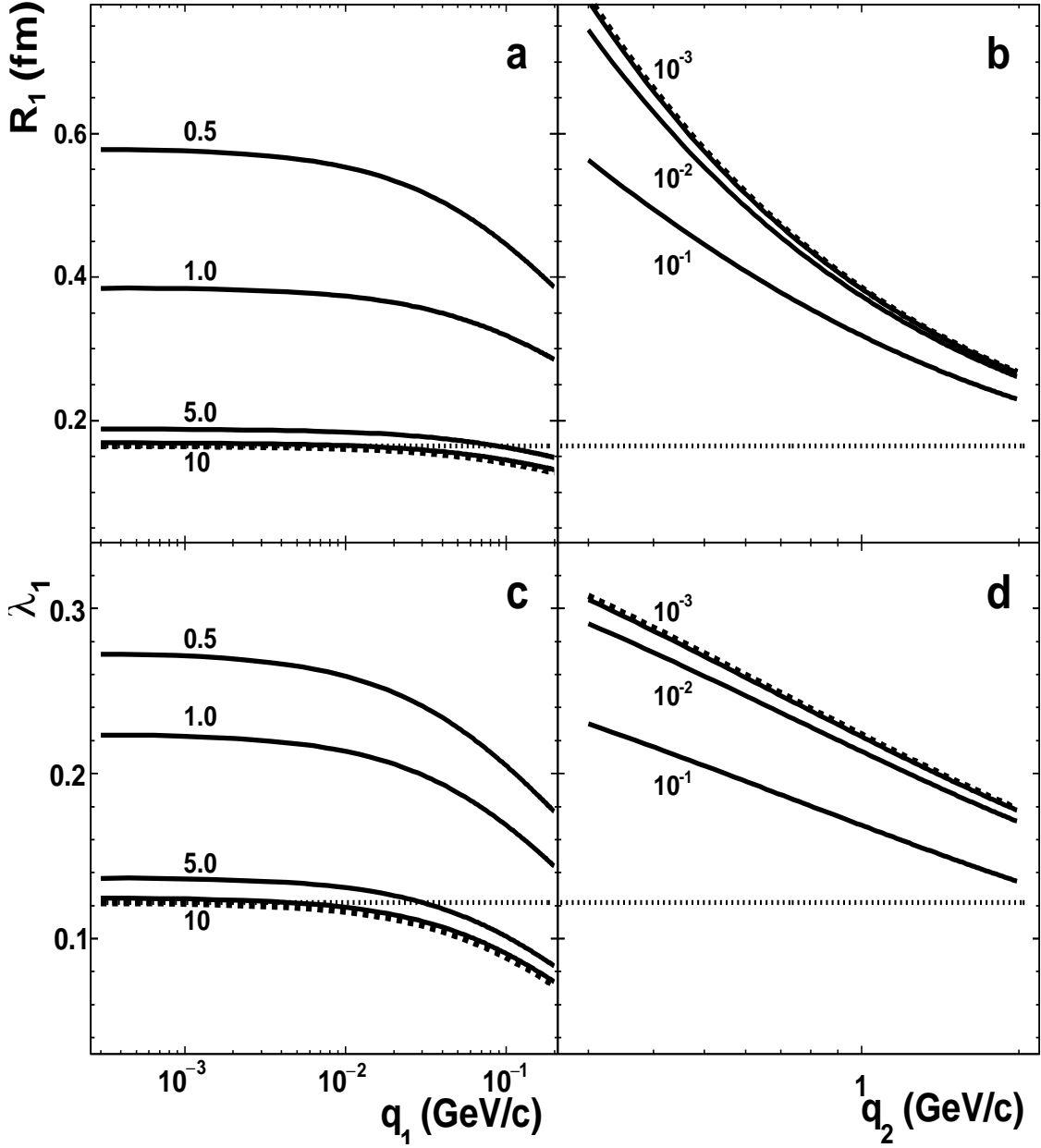}
\vspace*{8pt} \caption{Dependence of 1D BEC radius (a,b) and
strength of correlations (c,d) for centrally-symmetric L\'{e}vy
parametrization with $\alpha_{1}=1.5$ on low $q_{1}$ (a,c) and
high $q_{2}$ (b,d) limits of integration in the system of
equations (\ref{eq:2.1}) for fixed parameter values for second
centrally-symmetric L\'{e}vy parametrization: $\alpha_{2}=0.5$,
$\lambda_{2}=0.5$ and $R_{2}=1.5$ fm. The solid lines correspond
to the indicated values of the $q_{2}$ for $q_{1}$-dependence
(a,c) and to shown values of the $q_{1}$ for $q_{2}$-dependence
(b,d). The dashed lines correspond to the calculations with $q_{2}
\to \infty$ for $q_{1}$-dependence (a,c) and with $q_{1}=0$ for
$q_{2}$-dependence (b,d). The thin dotted lines are the ultimate
levels for $R_{1}$ (a,b) and $\lambda_{1}$ (c,d) calculated
with (\ref{eq:2.2}) for given values of the $\alpha_{1}$ and the
set of parameters $\{\alpha_{2},\lambda_{2},R_{2}\}$ for second
L\'{e}vy parametrization.} \label{fig:1}
\end{figure*}

Fig.\,\ref{fig:1} shows dependence of
1D BEC radius (a,b) and strength of correlations (c,d) for
centrally-symmetric L\'{e}vy parametrization $\Omega_{1}$ with
known $\alpha_{1}=1.5$ on low $q_{1}$ (a,c) and high $q_{2}$ (b,d)
limits of integration in the system (\ref{eq:2.1}) for set
$\{\alpha_{2},\lambda_{2},R_{2}\}=\{0.5,0.5,1.5~\mbox{fm}\}$ for
$\Omega_{2}$. The solid lines correspond to the indicated values
of the $q_{2}$ in GeV/$c$ for $q_{1}$-dependence (a,c) and to
shown values of the $q_{1}$ in GeV/$c$ for $q_{2}$-dependence
(b,d). Values of BEC parameters $\lambda_{1}$ and $R_{1}$ depend
strongly on the fixed second limit of integration ($q_{2(1)}$) for
both the $q_{1}$- (Fig.\,\ref{fig:1}a,c) and the
$q_{2}$-dependence (Fig.\,\ref{fig:1}b,d). The dashed lines
correspond to the results from the system (\ref{eq:2.1}) with
$q_{2} \to \infty$ for $q_{1}$-dependence (Fig.\,\ref{fig:1}a,c)
and with $q_{1}=0$ for $q_{2}$-dependence (Fig.\,\ref{fig:1}b,d).
As seen the curves for general case of (\ref{eq:2.1}) coincide
with dashed lines at $q_{2}=10$ GeV/$c$ for $q_{1}$-dependence
(Fig.\,\ref{fig:1}a,c) and at $q_{1}=10^{-3}$ GeV/$c$ for
$q_{2}$-dependence (Fig.\,\ref{fig:1}b,d). These values for
$q_{1}$ and especially for $q_{2}$ are far from the corresponding
limit in modern experimental CF. Therefore one should use the
system (\ref{eq:2.1}) with finite limits in an integrations in the
case of experimentally available $q$-ranges for two
parameterizations with $\Omega_{1}$ and $\Omega_{2}$. The thin
dotted lines demonstrate the ultimate levels for $R_{1}$
(Fig.\,\ref{fig:1}a,b) and $\lambda_{1}$ (Fig.\,\ref{fig:1}c,d)
calculated with (\ref{eq:2.2}) for given values of the
$\alpha_{1}$ and the set of parameters
$\{\alpha_{2},\lambda_{2},R_{2}\}$ for second L\'{e}vy
parametrization $\Omega_{2}$. One can use the simple relations
(\ref{eq:2.2}) for calculation $\{\lambda_{1},R_{1}\}$ at $q_{1}
\lesssim 10^{-2}$ GeV/$c$ and $q_{2} \gtrsim 10$ GeV/$c$
(Fig.\,\ref{fig:1}a,b) but as expected the values of BEC
parameters $\{\lambda_{1},R_{1}\}$ are far from the ultimate
levels at any $q_{1}$ for $q_{2}$-dependence in the range $q_{2}
\leq 2$ GeV/$c$ is considered in Fig.\,\ref{fig:1}b,d.

\begin{figure*}
\includegraphics[width=15.5cm,height=17.0cm]{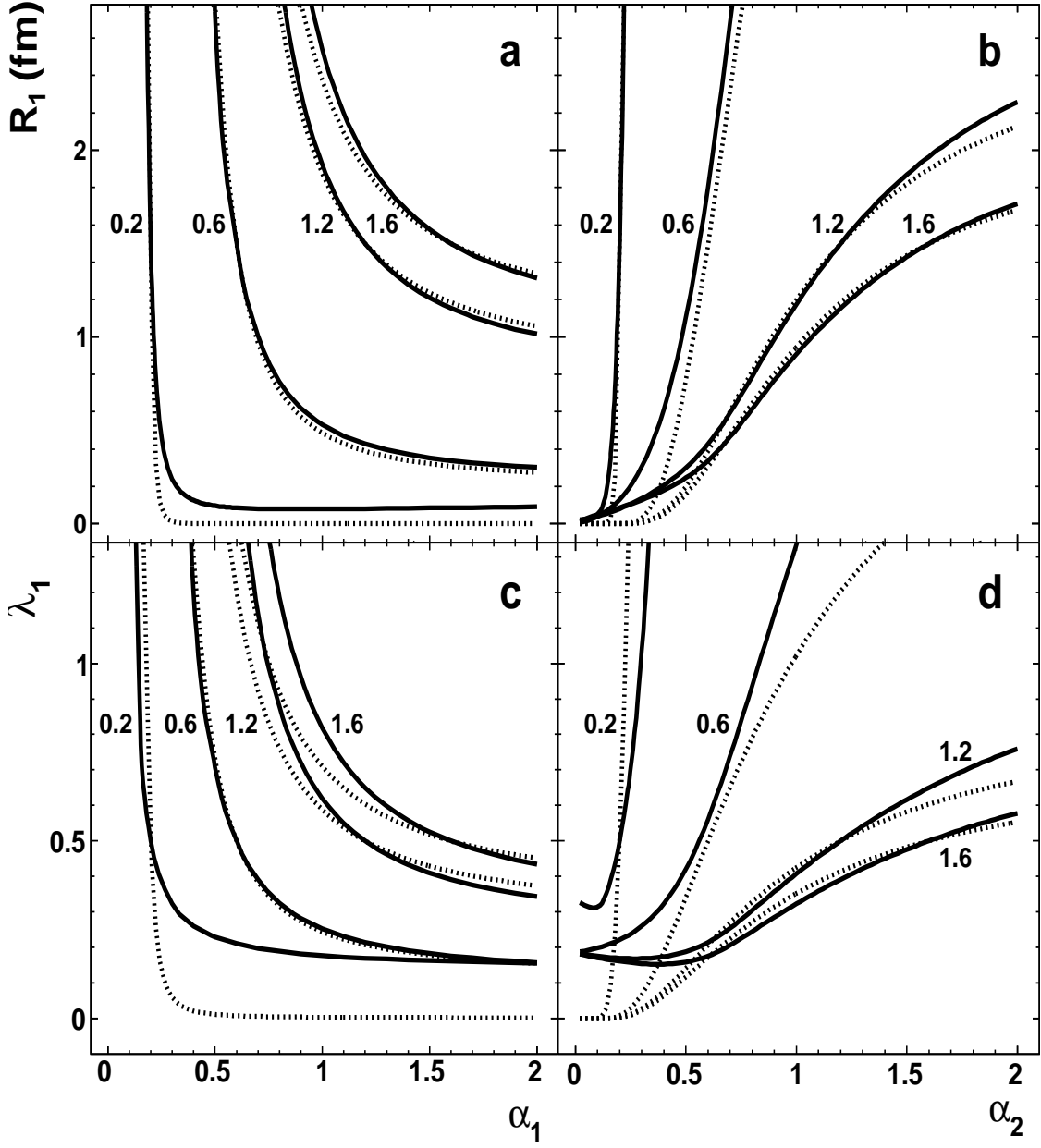}
\vspace*{8pt} \caption{Dependence of 1D BEC radius (a,b) and
strength of correlations (c,d) for centrally-symmetric L\'{e}vy
parametrization on $\alpha_{1}$ at fixed values of $\alpha_2$
(a,c) and on $\alpha_{2}$ at fixed values of $\alpha_{1}$ (b,d)
for given limits of integration in the system of equations
(\ref{eq:2.1}) $q_{1}=0.02$ GeV/$c$, $q_{2}=2.0$ GeV/$c$ and for
fixed values of the BEC parameters for second centrally-symmetric
L\'{e}vy parametrization: $\lambda_{2}=0.5$ and $R_{2}=1.5$ fm.
The solid lines correspond to the indicated values of the
$\alpha_{2}$ for $\alpha_{1}$-dependence (a,c) and to shown values
of the $\alpha_{1}$ for $\alpha_{2}$-dependence (b,d). The dotted
lines are the ultimate cases for $R_{1}$ (a,b) and
$\lambda_{1}$ (c,d) calculated with (\ref{eq:2.2}) for given
values of the $\alpha_{1}$ (a,c) or $\alpha_{2}$ (b,d) and the set
of BEC parameters $\{\lambda_{2},R_{2}\}$ for second L\'{e}vy
parametrization.} \label{fig:2}
\end{figure*}

In Fig.\,\ref{fig:2} dependence of 1D BEC radius (a,b) and
strength of correlations (c,d) is demonstrated for
centrally-symmetric L\'{e}vy parametrization with $\Omega_{1}$ on
$\alpha_{1}$ at fixed values of $\alpha_2$ (a,c) and on
$\alpha_{2}$ at fixed values of $\alpha_{1}$ (b,d) for given
limits of integration in the system of equations (\ref{eq:2.1})
$q_{1}=0.02$ GeV/$c$, $q_{2}=2.0$ GeV/$c$ and for certain values
of the BEC parameters for second centrally-symmetric L\'{e}vy
parametrization with $\Omega_{2}$: $\lambda_{2}=0.5$ and
$R_{2}=1.5$ fm. The solid lines correspond to the indicated values
of the $\alpha_{2}$ for $\alpha_{1}$-dependence (a,c) and to shown
values of the $\alpha_{1}$ for $\alpha_{2}$-dependence (b,d). The
values for limits of integration $q_{1}$ and $q_{2}$ in the system
(\ref{eq:2.1}) are similar to those used in modern experiments.
As seen dependencies of both BEC parameters on $\alpha_{i}$,
$i=1,2$ at fixed another L\'{e}vy index
$\left.\alpha_{j}\right|_{j \ne i}$, $j=1,2$ change very fast at
small value $\alpha_{j}=0.2$ in narrow range $\alpha_{i} \simeq
\alpha_{j}$. Such behavior is observed for both the results from
the system (\ref{eq:2.1}) and the estimations for $R_{1}$
(Fig.\,\ref{fig:2}a,b) and $\lambda_{1}$ (Fig.\,\ref{fig:2}c,d)
calculated with (\ref{eq:2.2}) for semi-infinite ranges for
integration and shown by dotted lines. The dependence
$R_{1}(\alpha_{1})$ shown in Fig.\,\ref{fig:2}a closes to the
analytic one calculated with help of (\ref{eq:2.2.b}) and
presented by dotted line for any $\alpha_{2}$ under study with
exception of the small value $\alpha_{2}=0.2$. For last case the
agreement is obtained in very narrow range $\alpha_{1} \approx
0.2$ between results from the system (\ref{eq:2.1}) and equation
(\ref{eq:2.2.b}). This feature maps clearly in corresponding
dependencies $R_{2}(\alpha_{2})$ shown in Fig.\,\ref{fig:2}b for
$\alpha_{1}=0.2$. For large $\alpha_{1}
> 1.0$ solid and dotted lines are close to each other in the range $\alpha_{2} >
0.5$ but agreement is poor significantly between results from the
system (\ref{eq:2.1}) and equation (\ref{eq:2.2.b}) for
$\alpha_{1}=0.6$ especially in domain $\alpha_{2} < 0.5$
(Fig.\,\ref{fig:2}b) taking into account the sharp behavior for
corresponding dependence $R_{1}(\alpha_{2})$. In general the
behavior of dependencies of $\lambda_{1}$ on $\alpha_{1}$
(Fig.\,\ref{fig:2}c) and $\alpha_{2}$ (Fig.\,\ref{fig:2}d) is
similar to the corresponding dependencies of 1D BEC radius
$R_{1}$. But the agreement is poor usually between the results
deduced from the system (\ref{eq:2.1}) and shown by the solid
lines and the estimations calculated based on the (\ref{eq:2.2.a})
and presented by the dotted lines. Therefore for values of limits
of integration $q_{1,2}$ under consideration the approximate
relations (\ref{eq:2.2}) should be used carefully for experimental
analysis of dependencies on L\'{e}vy indexes and last equations
can produce the reasonable estimations for BEC parameters for
ranges $\forall\,i=1,2: \alpha_{i} \gtrsim 1$ only.

As seen the mathematical formalism described above as well as the
results in Figs.\,\ref{fig:1}, \ref{fig:2} are quantitative basis
for choice of the applying of general equations (\ref{eq:2.1}) or
ultimate relations (\ref{eq:2.2}) in data analysis for given
experiment. Thus the method suggested in the paper is helpful for
experimental and phenomenological studies of BEC in various
processes at different parameterizations of CF $\mathbf{C}_{2}(q)$
corresponded to the centrally-symmetric L\'{e}vy source
distributions.

%#3
\section{Relations between BEC parameters in specific cases}\label{sec:3}

As seen in Fig.\,\ref{fig:2} both dependencies of the 1D BEC
radius $R_{1}$ (a,b) and the strength of correlations
$\lambda_{1}$ (c,d) on L\'{e}vy indexes $\alpha_{i}$, $i=1,2$ show
the weaker changing in the domain $\forall i=1, 2: \alpha_{i}
\gtrsim 1$ in comparison with the range of small values of
L\'{e}vy indexes. The region $\forall i=1, 2: \alpha_{i} \gtrsim
1$ includes in particular the specific cases of Cauchy and
Gaussian distributions for which corresponding parameterizations
of Bose\,--\,Einstein CF $\mathbf{C}_{2}(q)$ with $\alpha=1$ and
$\alpha=2$ are used mostly for experimental studies. Therefore
these certain views of $\Omega_{i}$ are studied in detail below.
Let $\Omega_{1} \equiv
\Omega_{\footnotesize{G}}=\Omega(2,\lambda_{\footnotesize{G}},R_{\footnotesize{G}})$
for Gaussian parametrization (\ref{eq:1.1}) and $\Omega_{2} \equiv
\Omega_{\footnotesize{C}}=\Omega(1,\lambda_{\footnotesize{C}},R_{\footnotesize{C}})$
for 1D approximation of experimental CF $\mathbf{C}_{2}(q)$ by
exponential function.

\subsection{Mathematical formalism}\label{subsec:3.1}
The relations (\ref{eq:2.2}) are valid at any values of indexes of
stability $0 < \alpha_{i} \leq 2$ in two centrally-symmetric
L\'{e}vy parameterizations with $\Omega_{i}$, $i=1,2$. If without
loss of generality the
$\{\lambda_{\footnotesize{C}},R_{\footnotesize{C}}\}$ are
considered as \emph{a priori} known and values of Gaussian BEC
parameters $\{\lambda_{\footnotesize{G}},R_{\footnotesize{G}}\}$
are supposed as desired then as expected the equations
(\ref{eq:2.2}) result in the ultimate relations:
\begin{subequations}
\begin{equation}
\lambda_{\footnotesize{G}}^{u}=2\lambda_{\footnotesize{C}}/\pi;
\label{eq:3.1.a}
\end{equation}
\vspace*{-0.5cm}
\begin{equation}
R_{\footnotesize{G}}^{\,u}=R_{\footnotesize{C}}/\sqrt{\pi}
\label{eq:3.1.b}
\end{equation}\label{eq:3.1}
\end{subequations}
\hspace*{-0.12cm}and vice versa. The relation (\ref{eq:3.1.a}) is
derived in \cite{Alexander-JPCS-675-022001-2016} for the first
time while the formula (\ref{eq:3.1.b}) for 1D BEC radii is
well-known.

The following relations can be obtained from general system
(\ref{eq:2.1}) for finite ranges of integrations and specific
values $\alpha_{1}=2$ and $\alpha_{2}=1$:
\begin{subequations}
\begin{equation}
\displaystyle
\frac{\lambda_{\footnotesize{G}}\sqrt{\pi}}{2R_{\footnotesize{G}}}
\sum_{j=1}^{N_{\footnotesize{G}}}[\mbox{erf}(z_{2,j,\footnotesize{G}})-\mbox{erf}(z_{1,j,\footnotesize{G}})]
= \frac{\lambda_{\footnotesize{C}}}{R_{\footnotesize{C}}}
\sum_{i=1}^{N_{\footnotesize{C}}}[\exp(-z_{1,i,\footnotesize{C}})-\exp(-z_{2,i,\footnotesize{C}})];
\label{eq:3.2.a}
\end{equation}
\vspace*{-0.5cm}
\begin{equation}
\displaystyle \frac{1}{R_{\footnotesize{G}}\sqrt{\pi}}
\frac{\sum_{j=1}^{N_{\footnotesize{G}}}[\exp(-z_{1,j,\footnotesize{G}}^{2})-\exp(-z_{2,j,\footnotesize{G}}^{2})]}
{\sum_{j=1}^{N_{\footnotesize{G}}}[\mbox{erf}(z_{2,j,\footnotesize{G}})-\mbox{erf}(z_{1,j,\footnotesize{G}})]}
= \frac{1}{R_{\footnotesize{C}}}
\frac{\sum_{i=1}^{N_{\footnotesize{C}}}[\exp(-z_{1,i,\footnotesize{C}})(1+z_{1,i,\footnotesize{C}})-\exp(-z_{2,i,\footnotesize{C}})(1+z_{2,i,\footnotesize{C}})]}
{\sum_{i=1}^{N_{\footnotesize{C}}}[\exp(-z_{1,i,\footnotesize{C}})-\exp(-z_{2,i,\footnotesize{C}})]}.
\label{eq:3.2.b}
\end{equation}\label{eq:3.2}
\end{subequations}
\hspace*{-0.12cm}Here $\mbox{erf}(x)=\frac{\textstyle
2}{\textstyle \sqrt{\pi}}\displaystyle\int_{0}^{x}\exp(-t^{2})dt$
is the error integral, $z_{1(2),i,\footnotesize{C(G)}} \equiv
q_{1(2),i,\footnotesize{C(G)}}R_{\footnotesize{C(G)}}$ are the
limits for integration over corresponding subranges for Cauchy
(Gaussian) distribution. The detailed study of all available experimental
results in strong interaction processes for 1D parametrization
(\ref{eq:1.1}) with $\alpha_{1}=2$ and $\alpha_{2}=1$ for
experimental CF $\mathbf{C}_{2}(q)$ shows that (i) the ranges of
integration for both the exponential and the Gaussian functions are equal;
(ii) usually, the range of integration is not divided into
subranges, in any case, such division is identical for both
functions under consideration and maximum value of the
$N_{\footnotesize{C(G)}}$ is equal 2 for experimental analyses.
Thus the general statement with regard of identity of integration
ranges for $\Omega_{1}$ and $\Omega_{2}$ is quite confirmed for
case of Cauchy and Gaussian distributions and $N_{\footnotesize{C(G)}}
\equiv N$ in the system (\ref{eq:3.2}).

Further simplification for the system of equations (\ref{eq:3.2})
depends on features of certain experiment; direction of
calculations $\Omega_{\footnotesize{C}} \rightleftarrows
\Omega_{\footnotesize{G}}$, i.e. what kind of a set of BEC parameters of the two, $\{\lambda_{\footnotesize{C}},R_{\footnotesize{C}}\}$
and $\{\lambda_{\footnotesize{G}},R_{\footnotesize{G}}\}$, it is regarded as
\emph{a priori} known, and which set is supposed as desired; and requirement on the accuracy level. For available
experimental data for BEC of charged pion pairs produced in strong
interaction processes (i) the accuracy for 1D BEC radius is better
usually than that for $\lambda$ parameter and (ii) the accuracy
for 1D BEC parameters in modern experiments is not better than $\sim 10^{-3}$ so far.
Thus one can assume the conservative accuracy level $\varepsilon =
5 \times 10^{-4}$. At present the most complex case with $N=2$ is
for analyses of proton-proton collisions at some LHC energies only
\cite{CMS-PRL-105-032001-2010,ATLAS-EPJC-75-466-2015}. For this
case all contributions are negligible from the subrange of $q$
values larger than the region of the influence of meson resonances
excluded from the experimental fits, i.e. all terms for $i=j=2$
can be omitted at given $\varepsilon$ and direction of calculation
from \emph{a priori} known Cauchy parameters to desired Gaussian
parameters $\{\lambda_{\footnotesize{C}},R_{\footnotesize{C}}\}
\to \{\lambda_{\footnotesize{G}},R_{\footnotesize{G}}\}$. But the
statement is wrong for opposite direction of calculation from
\emph{a priori} known Gaussian parameters to desired Gauchy
parameters $\{\lambda_{\footnotesize{G}},R_{\footnotesize{G}}\}
\to \{\lambda_{\footnotesize{C}},R_{\footnotesize{C}}\}$ at
$\varepsilon= 5 \times 10^{-4}$. Therefore the sum can be omitted
in the system (\ref{eq:3.2}) and equations can be re-written as
follow:
\begin{subequations}
\begin{equation}
\displaystyle
\frac{\lambda_{\footnotesize{G}}}{R_{\footnotesize{G}}\sqrt{\pi}}
[\mbox{erf}(z_{2,\footnotesize{G}})-\mbox{erf}(z_{1,\footnotesize{G}})]
= \frac{2\lambda_{\footnotesize{C}}}{\pi R_{\footnotesize{C}}}
[\exp(-z_{1,\footnotesize{C}})-\exp(-z_{2,\footnotesize{C}})];
\label{eq:3.3.a}
\end{equation}
\vspace*{-0.5cm}
\begin{equation}
\displaystyle \frac{1}{R_{\footnotesize{G}}\sqrt{\pi}}
\frac{\exp(-z_{1,\footnotesize{G}}^{2})-\exp(-z_{2,\footnotesize{G}}^{2})}
{\mbox{erf}(z_{2,\footnotesize{G}})-\mbox{erf}(z_{1,\footnotesize{G}})}
= \frac{1}{R_{\footnotesize{C}}}
\frac{\exp(-z_{1,\footnotesize{C}})(1+z_{1,\footnotesize{C}})-\exp(-z_{2,\footnotesize{C}})(1+z_{2,\footnotesize{C}})}
{\exp(-z_{1,\footnotesize{C}})-\exp(-z_{2,\footnotesize{C}})}
\label{eq:3.3.b}
\end{equation}\label{eq:3.3}
\end{subequations}
\hspace*{-0.12cm}for all experiments in the case of estimation of
unknown BEC parameters for Gaussian function based on the \emph{a
priori} known set of corresponding parameters for exponential function
and for all experiments with exception of $p+p$ collisions at
$\sqrt{s_{\footnotesize{NN}}}=0.9$, 2.36 and 7 TeV
\cite{CMS-PRL-105-032001-2010,ATLAS-EPJC-75-466-2015} in the case
of inverse problem. Account the relation $R_{\footnotesize{G}}
\leq R_{\footnotesize{C}}/\sqrt{\pi}$ and properties of the
functions $\exp(-x^{2}), \mbox{erf}(x)$ allows us to simplify Eqs.
(\ref{eq:3.3}) to the system
\begin{subequations}
\begin{equation}
\displaystyle
\frac{\lambda_{\footnotesize{G}}}{R_{\footnotesize{G}}\sqrt{\pi}}\,
\mbox{erfc}(z_{1,\footnotesize{G}}) =
\frac{2\lambda_{\footnotesize{C}}}{\pi R_{\footnotesize{C}}}
[\exp(-z_{1,\footnotesize{C}})-\exp(-z_{2,\footnotesize{C}})];
\label{eq:3.4.a}
\end{equation}
\vspace*{-0.5cm}
\begin{equation}
\displaystyle \frac{1}{R_{\footnotesize{G}}\sqrt{\pi}}
\frac{\exp(-z_{1,\footnotesize{G}}^{2})}
{\mbox{erfc}(z_{1,\footnotesize{G}})} =
\frac{1}{R_{\footnotesize{C}}}
\frac{\exp(-z_{1,\footnotesize{C}})(1+z_{1,\footnotesize{C}})-\exp(-z_{2,\footnotesize{C}})(1+z_{2,\footnotesize{C}})}
{\exp(-z_{1,\footnotesize{C}})-\exp(-z_{2,\footnotesize{C}})},
\label{eq:3.4.b}
\end{equation}\label{eq:3.4}
\end{subequations}
\hspace*{-0.12cm}where $\mbox{erfc}(x)=1-\mbox{erf}(x)$. The last
system of equations is valid for remain set of experimental
results with exception of the WA98 data
\cite{WA98-EPJC-16-445-2000} at given $\varepsilon$. Also the
transition from system (\ref{eq:3.3}) to simpler equations
(\ref{eq:3.4}) is not valid for CPLEAR data
\cite{CPLEAR-ZPC-63-541-1994} for direction of calculation from
\emph{a priori} known Gaussian parameters to desired Gauchy
parameters $\{\lambda_{\footnotesize{G}},R_{\footnotesize{G}}\}
\to \{\lambda_{\footnotesize{C}},R_{\footnotesize{C}}\}$ at
$\varepsilon= 5 \times 10^{-4}$. The simplest view of the system
of equations (\ref{eq:3.4})
\begin{subequations}
\begin{equation}
\displaystyle
\frac{\lambda_{\footnotesize{G}}}{R_{\footnotesize{G}}\sqrt{\pi}}\,
\mbox{erfc}(z_{1,\footnotesize{G}}) =
\frac{2\lambda_{\footnotesize{C}}}{\pi R_{\footnotesize{C}}}
\exp(-z_{1,\footnotesize{C}}); \label{eq:3.5.a}
\end{equation}
\vspace*{-0.5cm}
\begin{equation}
\displaystyle \frac{1}{R_{\footnotesize{G}}\sqrt{\pi}}
\frac{\exp(-z_{1,\footnotesize{G}}^{2})}
{\mbox{erfc}(z_{1,\footnotesize{G}})} =
\frac{1+z_{1,\footnotesize{C}}}{R_{\footnotesize{C}}}
\label{eq:3.5.b}
\end{equation}\label{eq:3.5}
\end{subequations}
\hspace*{-0.12cm}corresponds to the range of integration
$[z_{1,\footnotesize{C(G)}},\infty)$ and can be used for
experimental results from ALICE \cite{ALICE-PRD-82-052001-2010},
CMS
\cite{CMS-PRL-105-032001-2010,CMS-JHEP-0511-029-2011,Padula-WPCF-2014,Sikler-arXiv-1411.6609-2014}
with exception of the collision energy $\sqrt{s_{\footnotesize{NN}}}=2.36$ TeV
\cite{CMS-PRL-105-032001-2010} in the case of proton-proton
collisions and WA80 \cite{Albrecht-ZPC-53-225-1992} for asymmetric
nucleus-nucleus collisions $\mbox{O}+\mbox{C}$,
$\mbox{O}+\mbox{Cu}$. On the other hand, the using of the range of
integration $[0.0,z_{2,\footnotesize{C(G)}}]$ allows the
derivation from the Eqs. (\ref{eq:3.4}) the following system:
\begin{subequations}
\begin{equation}
\displaystyle
\frac{\lambda_{\footnotesize{G}}}{R_{\footnotesize{G}}\sqrt{\pi}}\,
\mbox{erf}(z_{2,\footnotesize{G}}) =
\frac{2\lambda_{\footnotesize{C}}}{\pi R_{\footnotesize{C}}}
[1-\exp(-z_{2,\footnotesize{C}})]; \label{eq:3.6.a}
\end{equation}
\vspace*{-0.5cm}
\begin{equation}
\displaystyle \frac{1}{R_{\footnotesize{G}}\sqrt{\pi}}
\frac{1-\exp(-z_{2,\footnotesize{G}}^{2})}
{\mbox{erf}(z_{2,\footnotesize{G}})} =
\frac{1}{R_{\footnotesize{C}}}
\frac{1-\exp(-z_{2,\footnotesize{C}})(1+z_{2,\footnotesize{C}})}{1-\exp(-z_{2,\footnotesize{C}})}
\label{eq:3.6.b}
\end{equation}\label{eq:3.6}
\end{subequations}
\hspace*{-0.12cm}As expected one can get the ultimate relations
(\ref{eq:3.1}) from the any systems of equations (\ref{eq:3.5}) or
(\ref{eq:3.6}) at $q_{1} \to 0$ or $q_{2} \to \infty$
respectively. Therefore the system (\ref{eq:3.5}) can be replaced by ultimate system of equations
(\ref{eq:3.1}) with some accuracy $\varepsilon'$ for finite range
of $q$ if $q_{1} \leq q_{1}^{h}$ and $q_{2}$ value is large enough
to consider this value as $q_{2} \to \infty$. Similarly, the system
(\ref{eq:3.6}) can be replaced by ultimate system of equations
(\ref{eq:3.1}) with some accuracy $\varepsilon'$ for finite range
of $q$ if $q_{2} \geq q_{2}^{l}$ and $q_{1}$ is small
enough to consider it as $q_{1} \to 0$ for Eqs. (\ref{eq:3.6}).
The
high / low boundary values $q_{1}^{h}$ / $q_{2}^{l}$ for variables
$q_{1}$ / $q_{2}$ are dominated by assigned value of accuracy. For
instance, at $\varepsilon' = 10^{-2}$ the ultimate system of
equations (\ref{eq:3.1}) is valid for $q_{1} \lesssim 2 \times
10^{-3}R_{\footnotesize{C(G)}}$ or $q_{2} \gtrsim
1.3R_{\footnotesize{C(G)}}$, i.e. $q_{1} \lesssim q_{1}^{h}=2-4$
MeV/$c$ or $q_{2} \gtrsim q_{2}^{l}=1.3-2.6$ GeV/$c$ for
proton-proton collisions. The derived estimations are close to the
values of $q$ variable which used in present experimental analyses
of BEC correlations.

These qualitative estimations are confirmed by quantitative
analysis below for the $q_{1}$-, $q_{2}$-dependencies of the Gaussian
parameters $\lambda_{\footnotesize{G}}$ and $R_{\footnotesize{G}}$
derived for some assigned values of the corresponding BEC
parameters for exponential function $\lambda_{\footnotesize{C}}$,
$R_{\footnotesize{C}}$ and vice versa.

\subsection{Dependence on q for desired Cauchy / Gaussian
parameters}\label{subsec:3.2}
\begin{figure*}
\includegraphics[width=15.5cm,height=17.0cm]{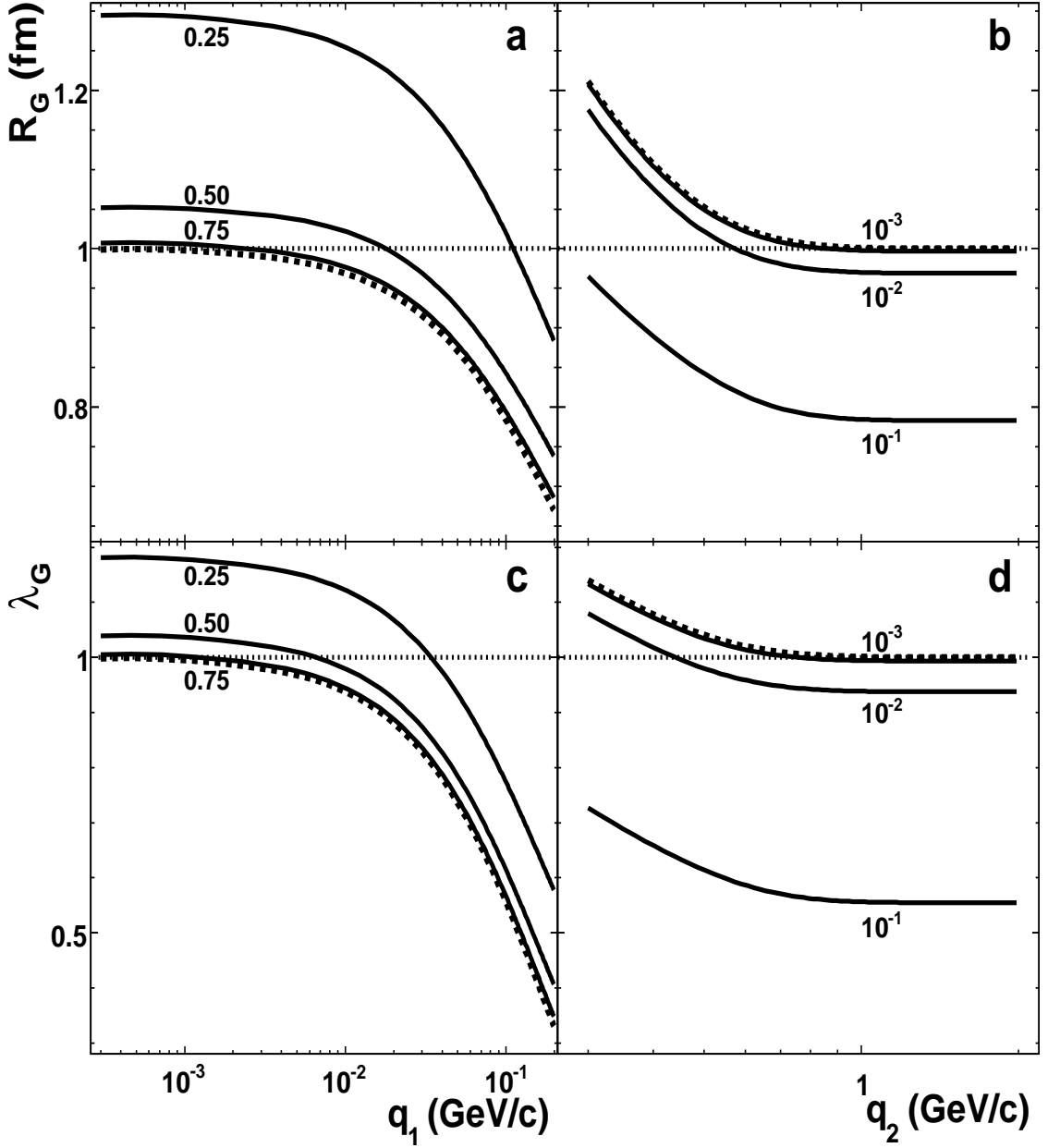}
\vspace*{8pt} \caption{Dependence of 1D BEC radius (a,b) and
strength of correlations (c,d) for Gaussian parametrization on low
$q_{1}$ (a,c) and high $q_{2}$ (b,d) limits of integration in the
system of equations (\ref{eq:3.3}) for fixed values of the
parameters for exponential parametrization:
$\lambda_{\footnotesize{C}}=\pi / 2$ and
$R_{\footnotesize{C}}=\sqrt{\pi}$ fm. The solid lines correspond
to the indicated values of the $q_{2}$ for $q_{1}$-dependence
(a,c) and to shown values of the $q_{1}$ for $q_{2}$-dependence
(b,d). The dashed lines correspond to the calculations based on
the system (\ref{eq:3.5}) for $q_{1}$-dependence (a,c) and on the
system (\ref{eq:3.6}) for $q_{2}$-dependence (b,d). The thin
dotted lines are the ultimate levels $R_{\footnotesize{G}}=1.0$ fm
(a,b) and $\lambda_{\footnotesize{G}}=1.0$ (c,d) calculated with
help of (\ref{eq:3.1}) for given values of the set of Cauchy
parameters $\{\lambda_{\footnotesize{C}},R_{\footnotesize{C}}\}$.}
\label{fig:3}
\end{figure*}

\begin{figure*}
\includegraphics[width=15.5cm,height=17.0cm]{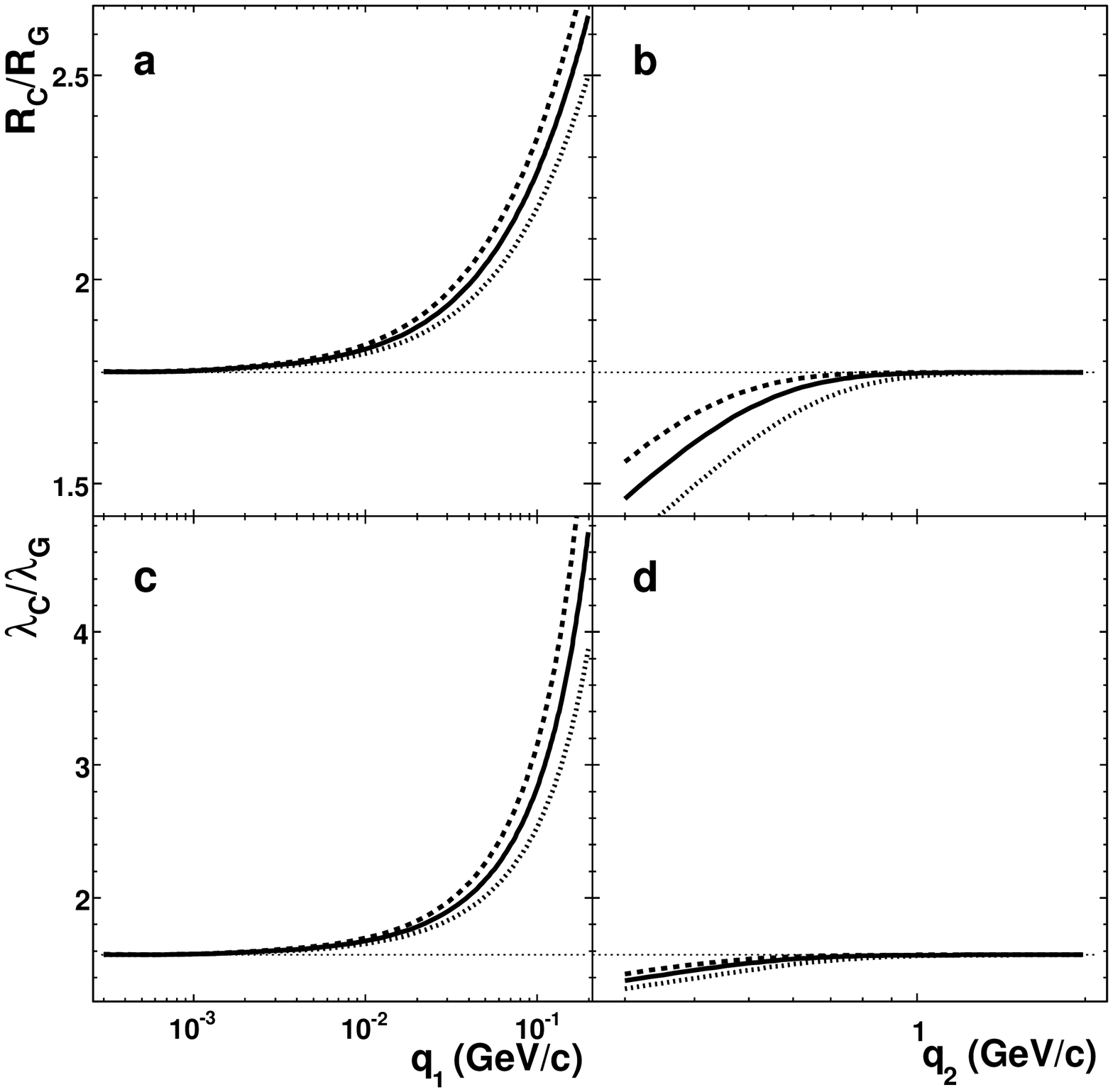}
\vspace*{8pt} \caption{Dependence of relative 1D BEC radius (a,b)
and strength of correlations (c,d) on $q_{1}$ (a,c) and $q_{2}$
(b,d) for various fixed values of the parameters for exponential parametrization. The calculations are made for simpler range of
integration $[z_{1,\footnotesize{C(G)}}, \infty)$ with help of
system (\ref{eq:3.5}) for $q_{1}$-dependence (a,c) and for
$[0.0,z_{2,\footnotesize{C(G)}}]$ with system (\ref{eq:3.6}) for
$q_{2}$-dependence (b,d) respectively. The dashed lines correspond
to the $\lambda_{\footnotesize{C}}=0.6 \pi$,
$R_{\footnotesize{C}}=1.2\sqrt{\pi}$ fm; solid lines --
$\lambda_{\footnotesize{C}}=0.5\pi$,
$R_{\footnotesize{C}}=\sqrt{\pi}$ fm; dotted lines --
$\lambda_{\footnotesize{C}}=0.4 \pi$,
$R_{\footnotesize{C}}=0.8\sqrt{\pi}$ fm. The thin dotted lines are
the ultimate levels
$R_{\footnotesize{C}}/R_{\footnotesize{G}}=\sqrt{\pi}$ (a,b) and
$\lambda_{\footnotesize{C}}/\lambda_{\footnotesize{G}}=\pi/2$
(c,d) corresponded to the system (\ref{eq:3.1}).} \label{fig:4}
\end{figure*}

For Fig.\,\ref{fig:3}, \ref{fig:4}
the $\Omega_{\footnotesize{C}}$ is considered as \emph{a priori}
known and set of BEC parameters
$\{\lambda_{\footnotesize{G}},R_{\footnotesize{G}}\}$ are
studied for Gaussian parametrization (\ref{eq:1.1}). The
Fig.\,\ref{fig:3} shows the $q_{1}$- and $q_{2}$-dependence of 1D
BEC radius (Fig.\,\ref{fig:3}a,b) and strength of correlations
(Fig. \ref{fig:3}c,d) for parametrization (\ref{eq:1.1}) with
Gaussian function $\Omega_{\footnotesize{G}}$ at fixed values
$\lambda_{\footnotesize{C}}=\pi / 2$ and
$R_{\footnotesize{C}}=\sqrt{\pi}$. As seen the both Gaussian
parameters show the similar behavior with changing the integration
limits, namely $\lambda_{\footnotesize{G}}$ and
$R_{\footnotesize{G}}$ growth with decreasing of the $q_{1,2}$ at
fixed another limit of integration. The curves
$\lambda_{\footnotesize{G}}(q_{1})$, $R_{\footnotesize{G}}(q_{1})$
approach to the asymptotic dashed lines calculated with help of
the system (\ref{eq:3.5}) with increasing of the $q_{2}$. The
similar situation is observed in Fig.\,\ref{fig:3}b,d for curves
$R_{\footnotesize{G}}(q_{2})$, $\lambda_{\footnotesize{G}}(q_{2})$
and asymptotic dashed lines calculated with help of the system
(\ref{eq:3.6}) with decreasing of the $q_{1}$. As seen the
asymptotic lines are achieved at $q_{1} \lesssim 1$ MeV/$c$
(Fig.\,\ref{fig:3}b,d) and $q_{2} \gtrsim 0.8$ GeV/$c$
(Fig.\,\ref{fig:3}a,c). Furthermore the ultimate values of the
Gaussian BEC parameters $\lambda_{\footnotesize{G}}^{u}$ and
$R_{\footnotesize{G}}^{\,u}$ are valid with good accuracy for
$q_{1} < 2.0$ MeV/$c$ and $q_{2}
> 1.0$ GeV/$c$. The last ranges are in the good agreement with
qualitative estimations for proton-proton collisions obtained
above. It should be emphasized that for specific case of exponential
($\Omega_{\footnotesize{C}}$) and Gaussian
($\Omega_{\footnotesize{G}}$) functions the asymptotic
$q_{1}$-dependence is achieved for both the 1D BEC radius
(Fig.\,\ref{fig:3}a) and the strength of correlations
(Fig.\,\ref{fig:3}c) at $q_{2}$ which is much smaller than that
for case of two some another centrally-symmetric L\'{e}vy
parameterizations (Fig.\,\ref{fig:3}a,c). This $q_{2}$ value for
case of $\Omega_{\footnotesize{C}}$ and
$\Omega_{\footnotesize{G}}$ is similar to those used in analyses of
experimental CF $\mathbf{C}_{2}(q)$. In Fig.\,\ref{fig:4} the
dependencies of relative BEC parameters, namely
$R_{\footnotesize{C}}/R_{\footnotesize{G}}$ (a,b) and
$\lambda_{\footnotesize{C}}/\lambda_{\footnotesize{G}}$ (c,d), on
$q_{1}$ (a,c) and $q_{2}$ (b,d) are presented for various assigned
values of parameters for Cauchy distribution. The curves are
calculated with the simpler system of equations (\ref{eq:3.5}) for
$q_{1}$-dependence (Figs.\,\ref{fig:4}a,c) and system
(\ref{eq:3.6}) for $q_{2}$-dependence (Fig.\,\ref{fig:4}b,d)
respectively. As seen the larger values of Cauchy parameters lead
to the larger values of relative BEC parameters. The
$q_{1}$-dependence of relative BEC parameters growth faster with
increasing of the input values of Cauchy parameters
(Fig.\,\ref{fig:4}a,c). On the contrary the decrease of the
$q_{2}$-dependence of the
$R_{\footnotesize{C}}/R_{\footnotesize{G}}$ (Fig.\,\ref{fig:4}b)
and $\lambda_{\footnotesize{C}}/\lambda_{\footnotesize{G}}$
(Fig.\,\ref{fig:4}d) is slower with increasing of the input values
of the $\{\lambda_{\footnotesize{C}},R_{\footnotesize{C}}\}$. As
expected the ultimate levels
$R_{\footnotesize{C}}/R_{\footnotesize{G}}=\sqrt{\pi}$
(Fig.\,\ref{fig:4}a,b) and
$\lambda_{\footnotesize{C}}/\lambda_{\footnotesize{G}}=\pi/2$
(Fig.\,\ref{fig:4}c,d) shown by thin dotted lines are valid for
the same ranges of $q_{1}$ and $q_{2}$ as estimated above for
Fig.\,\ref{fig:3}.

\begin{figure*}
\includegraphics[width=15.5cm,height=17.0cm]{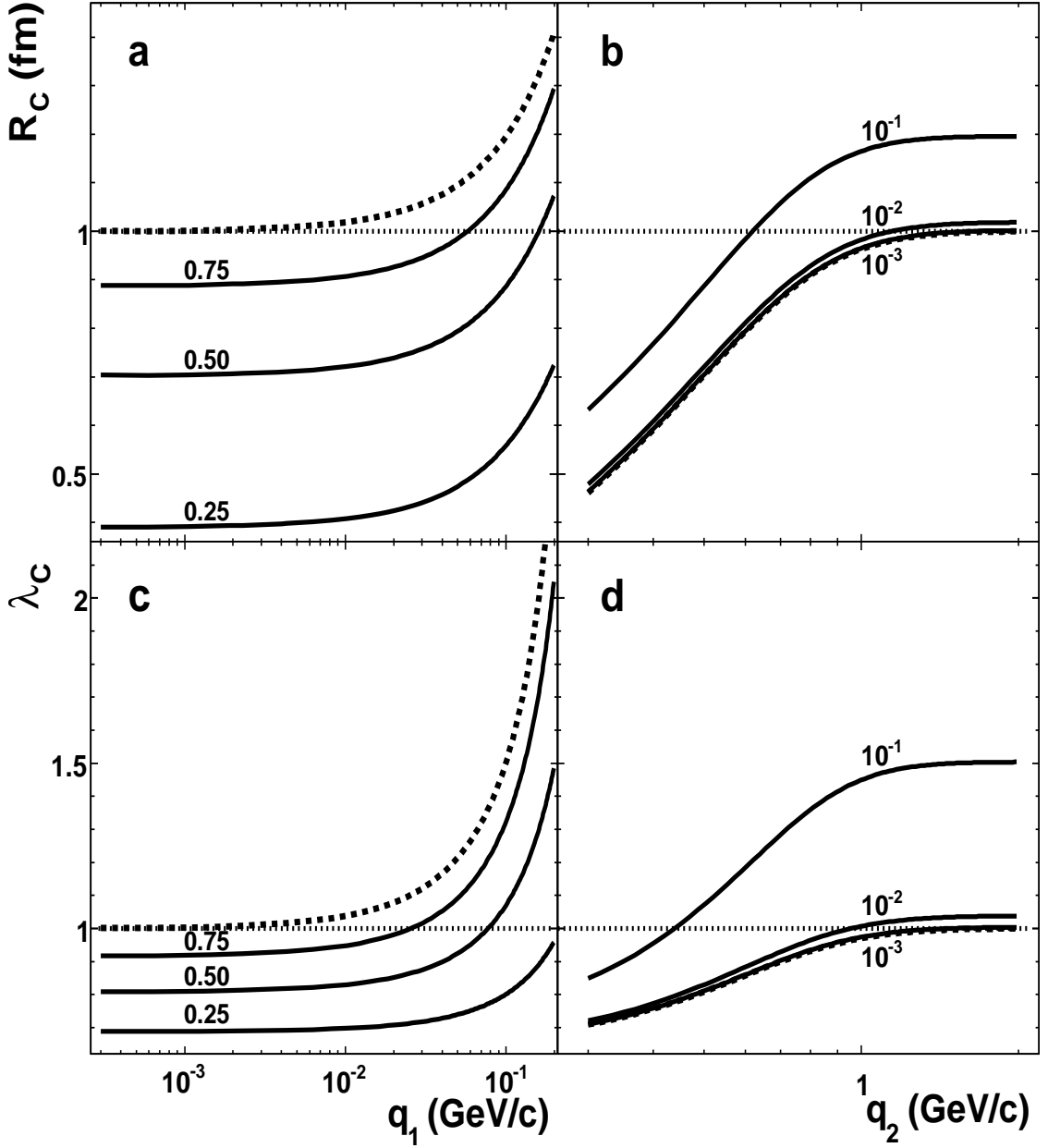}
\vspace*{8pt} \caption{Dependence of 1D BEC radius (a,b) and
strength of correlations (c,d) for Cauchy distribution on low
$q_{1}$ (a,c) and high $q_{2}$ (b,d) limits of integration in the
system of equations (\ref{eq:3.3}) for fixed values of the
parameters for Gaussian parametrization:
$\lambda_{\footnotesize{G}}=2/\pi$ and
$R_{\footnotesize{G}}=1/\sqrt{\pi}$ fm. The solid lines correspond
to the indicated values of the $q_{2}$ for $q_{1}$-dependence
(a,c) and to shown values of the $q_{1}$ for $q_{2}$-dependence
(b,d). The dashed lines correspond to the calculations based on
the system (\ref{eq:3.5}) for $q_{1}$-dependence (a,c) and on the
system (\ref{eq:3.6}) for $q_{2}$-dependence (b,d). The thin
dotted lines are the ultimate levels $R_{\footnotesize{C}}=1.0$ fm
(a,b) and $\lambda_{\footnotesize{G}}=1.0$ (c,d) calculated with
help of (\ref{eq:3.1}) for given values of the set of Gaussian
parameters $\{\lambda_{\footnotesize{G}},R_{\footnotesize{G}}\}$.}
\label{fig:5}
\end{figure*}

\begin{figure*}
\includegraphics[width=15.5cm,height=17.0cm]{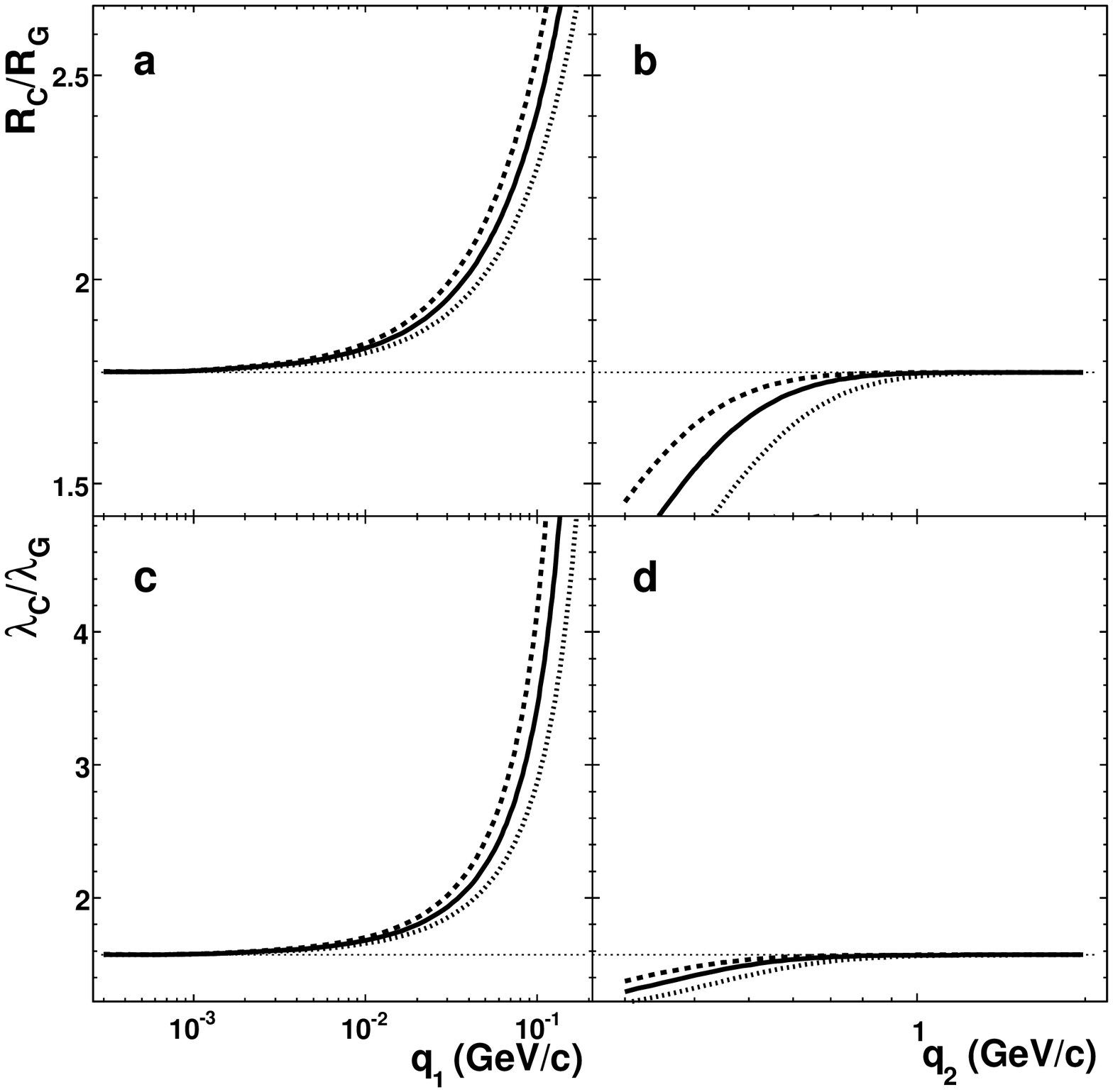}
\vspace*{8pt} \caption{Dependence of relative 1D BEC radius (a,b)
and strength of correlations (c,d) on $q_{1}$ (a,c) and $q_{2}$
(b,d) for various fixed values of the parameters for Gaussian
parametrization. The calculations are made for simpler range of
integration $[z_{1,\footnotesize{C(G)}}, \infty)$ with help of
system (\ref{eq:3.5}) for $q_{1}$-dependence (a,c) and for
$[0.0,z_{2,\footnotesize{C(G)}}]$ with system (\ref{eq:3.6}) for
$q_{2}$-dependence (b,d) respectively. The dashed lines correspond
to the $\lambda_{\footnotesize{G}}=1.2$,
$R_{\footnotesize{G}}=1.2$ fm; solid lines --
$\lambda_{\footnotesize{G}}=1.0$, $R_{\footnotesize{G}}=1.0$ fm;
dotted lines -- $\lambda_{\footnotesize{G}}=0.8$,
$R_{\footnotesize{G}}=0.8$ fm. The thin dotted lines are the
ultimate levels
$R_{\footnotesize{C}}/R_{\footnotesize{G}}=\sqrt{\pi}$ (a,b) and
$\lambda_{\footnotesize{C}}/\lambda_{\footnotesize{G}}=\pi/2$
(c,d) corresponded to the system (\ref{eq:3.1}).} \label{fig:6}
\end{figure*}

Fig.\,\ref{fig:5}, \ref{fig:6} shows results for opposite
direction of calculations, i.e $\Omega_{\footnotesize{G}}$ is
supposed \emph{a priori} known and BEC parameters
$\{\lambda_{\footnotesize{C}},R_{\footnotesize{C}}\}$ for exponential parametrization of CF $\mathbf{C}_{2}(q)$  are derived. Fig.\,\ref{fig:5} shows the $q_{1}$- and
$q_{2}$-dependence of 1D BEC radius (Fig.\,\ref{fig:5}a,b) and
strength of correlations (Fig.\,\ref{fig:5}c,d) for exponential
function $\Omega_{\footnotesize{C}}$ at fixed values
$\lambda_{\footnotesize{G}}=2 / \pi$ and $R_{\footnotesize{G}}=1 /
\sqrt{\pi}$. The $\forall\,i=1,2: q_{i}$-dependencies show the
opposite behavior for desired parameters of Cauchy source function
$R_{\footnotesize{C}}$ (Fig.\,\ref{fig:5}a,b) and
$\lambda_{\footnotesize{C}}$ (Fig.\,\ref{fig:5}c,d) with respect
to the corresponding dependencies presented in Fig.\,\ref{fig:3}
above for another direction of calculation
$\{\lambda_{\footnotesize{C}},R_{\footnotesize{C}}\} \to
\{\lambda_{\footnotesize{G}},R_{\footnotesize{G}}\}$. These
differences are seen in domain of relatively large $q_{1} \gtrsim
10^{-2}$ GeV/$c$ for $q_{1}$-dependence and at relatively small
$q_{2} \lesssim 0.7$ GeV/$c$ for $q_{2}$-dependence of BEC
parameters. Furthermore the $q_{1}$-dependence for parameters from
set for $\Omega_{\footnotesize{C}}$ (Fig.\,\ref{fig:5}a,c)
approaches to the constant at $q_{1} \to 0$ faster noticeably than
that for $R_{\footnotesize{G}}$ (Fig.\,\ref{fig:3}a) and
$\lambda_{\footnotesize{G}}$ (Fig.\,\ref{fig:3}c). The opposite
situation is observed for achievement of constants by
$q_{2}$-dependence at $q_{2} \to \infty$. It should be noted that
dependencies $R_{\footnotesize{C}}(q_{1})$ and
$\lambda_{\footnotesize{C}}(q_{1})$ approach to its asymptotic
curves calculated with help of the system (\ref{eq:3.5}) and shown
by dashed lines in Fig.\,\ref{fig:5}a,c slower than corresponding
dependencies for desired Gaussian parameters in Fig.\,\ref{fig:3}a,c.
As consequence the $R_{\footnotesize{C}}(q_{1})$ and
$\lambda_{\footnotesize{C}}(q_{1})$ will achieve the asymptotic
curves at higher $q_{2}$ than that for Fig.\,\ref{fig:3}a,c. The
asymptotic value of $q_{1} \simeq 10^{-3}$ GeV/$c$ is the same for
$q_{2}$-dependence for both directions of calculations
$\{\lambda_{\footnotesize{C}},R_{\footnotesize{C}}\}
\rightleftarrows
\{\lambda_{\footnotesize{G}},R_{\footnotesize{G}}\}$.
Fig.\,\ref{fig:6} demonstrates the dependence of relative BEC
parameters, namely $R_{\footnotesize{C}}/R_{\footnotesize{G}}$
(a,b) and $\lambda_{\footnotesize{C}}/\lambda_{\footnotesize{G}}$
(c,d), on $q_{1}$ (a,c) and $q_{2}$ (b,d) for various assigned
values of parameters for Gaussian parametrization. The simpler system
of equations (\ref{eq:3.5}) is used for calculation of
$q_{1}$-dependencies in Figs.\,\ref{fig:6}a,c and curves on
$q_{2}$ (Fig.\,\ref{fig:6}b,d) are derived with help of the system
(\ref{eq:3.6}). In general $\forall\,i=1,2: q_{i}$-dependencies
show similar behavior for corresponding relative 1D BEC parameters
in both cases the Fig.\,\ref{fig:4} and the Fig.\,\ref{fig:6} with
some faster changing of $q_{i}$-dependencies in the second case
than that for the first one in domain of relatively large $q_{1}
\gtrsim 10^{-2}$ GeV/$c$ for $q_{1}$-dependence and at relatively
small $q_{2} \lesssim 0.7$ GeV/$c$ for $q_{2}$-dependence of
$R_{\footnotesize{C}}/R_{\footnotesize{G}}$ and
$\lambda_{\footnotesize{C}}/\lambda_{\footnotesize{G}}$.

Simultaneous consideration of available 1D BEC data analyses for
strong interaction processes and Fig.\,\ref{fig:3}, \ref{fig:5}
allows the assertion that ultimate relations (\ref{eq:3.1}) are
not acceptable with reasonable accuracy for most of experimental
results with exponential / Gaussian parametrization (\ref{eq:1.1}) of 1D
CF $\mathbf{C}_{2}(q)$. As seen from Fig.\,\ref{fig:4},
\ref{fig:6} even the asymptotic values of relative 1D BEC
parameters
$\{\lambda_{\footnotesize{C}}/\lambda_{\footnotesize{G}},
R_{\footnotesize{C}}/R_{\footnotesize{G}}\}$ can differ up to
several times from ultimate values calculated with help of the
system of equations (\ref{eq:3.1}) in some domains of $q_{1}$ and
$q_{2}$ variables. Therefore Figs.\,\ref{fig:3}\,--\,\ref{fig:6}
confirm the conclusion formulated above for case of two general
view centrally-symmetric L\'{e}vy parameterizations, namely, for
desired 1D BEC parameters the finite values for limits of
integrations can lead to the significant difference between values
of BEC observables calculated on exact equations and asymptotic /
ultimate values calculated on simpler relations.

It should be emphasized the results of the present paper shown in
Figs.\,\ref{fig:3}\,--\,\ref{fig:6} are useful for experimental
data analysis as well as for phenomenological studies because it
allow, in particular, the quantitative choice between systems of
equations (\ref{eq:3.1}) -- (\ref{eq:3.6}) for estimations of 1D
BEC parameters for specific cases of centrally-symmetric L\'{e}vy
parametrization (\ref{eq:1.1}) at $\alpha=1, 2$ depending on some
features in given experiment.

%#4
\section{Comparison with experimental results}\label{sec:4}

Database is created for 1D BEC results for identical charged pions
produced in strong interaction processes in order to verify the
mathematical formalisms suggested above. This database is shown in
the Appendix\,\ref{sec:6} and it is used as input for calculations
below. Experimental results for strength of correlations and 1D
source radius are considered for all types of the processes,
centrally-symmetric L\'{e}vy parameterizations (\ref{eq:1.1}) and
for total available energy range in the paper. The results for
most central nucleus-nucleus collisions are usually included in
the database because these collisions are used for studying of new
features of final-state matter
\cite{Okorokov-AHEP-2015-790646-2015}. The dependence of 1D BEC
parameters on the outgoing charged particle multiplicity,
$N_{ch}$, is wide studied for $p+p$ and $\bar{p}+p$ collisions at
least. Therefore the additional separation is made on experimental
1D BEC values deduced for minimum bias and for high multiplicity
event classes sometimes\footnote{This separation will be
stipulated additionally if experimental 1D BEC results are
available for various multiplicity event classes in $p+p$,
$\bar{p}+p$ collisions.}. This consideration seems important for
both the additional verification of mathematics above and the more
careful comparison with nucleus-nucleus results. As seen the
additional information is required about experimental $q$ ranges
for systems (\ref{eq:2.1}), (\ref{eq:3.2}) in comparison with the
ultimate relations (\ref{eq:2.2}) and (\ref{eq:3.1}). Therefore
experimental $q$ ranges are estimated based on the available
published data. In this Section in Tables
\ref{tab:4.1}--\ref{tab:4.3} the statistical errors are
shown first, available systematic uncertainties -- second, unless
otherwise specifically indicated; Ithe types of uncertainties
(statistical / total, symmetric / asymmetric) is chosen just the
same as well as input parameters for the sake of simplicity.

\subsection{Relations between parameters for Cauchy / Gaussian
distribution and L\'{e}vy one}\label{subsec:4.1} The general
system of equations (\ref{eq:2.1}) allow us to estimate the 1D BEC
parameters for exponential / Gaussian function
$\Omega_{\footnotesize{C(G)}}$ based on the \emph{a priori} known
parameter values for $\Omega_{\footnotesize{L}} \equiv
\Omega(\alpha_{\footnotesize{L}},\lambda_{\footnotesize{L}},R_{\footnotesize{L}})$
corresponded to general view of centrally-symmetric L\'{e}vy
distribution and vice versa.

\textbf{1.~Direction of calculations $\Omega_{\footnotesize{L}}
\to \Omega_{\footnotesize{C/G}}$.}~~ The sets
$\{\lambda_{\footnotesize{C}},R_{\footnotesize{C}}\}$ and
$\{\lambda_{\footnotesize{G}},R_{\footnotesize{G}}\}$ are
estimated for experimentally known $\Omega_{\footnotesize{L}}$
\cite{CMS-PRL-105-032001-2010,CMS-JHEP-0511-029-2011,ATLAS-Astalos-PhD2015}
and finite $q$-ranges with help of (\ref{eq:2.1}). The estimations
are shown in Table \ref{tab:4.1} together with the available
experimental results and the data for Cauchy distribution are
shown on the first line, for Gaussian parametrization -- on the
second line for certain experiment at given energy. As seen
estimations for strength of correlations and 1D radius calculated
with help of (\ref{eq:2.1}) agree with experimental values within
errors for both the Cauchy and the Gaussian distributions for
particle emission points at all energies under study. One can note
that estimation for $\lambda_{\footnotesize{G}}$ coincides with
experimental values within total errors only at
$\sqrt{s_{\footnotesize{NN}}}=7$ TeV. Nevertheless the general
system of equations (\ref{eq:2.1}) provides rather well
estimations of 1D BEC parameters for both the Cauchy and the
Gaussian distributions. The possibility is considered for
application of ultimate relations (\ref{eq:2.2}) for the
experimental data under study. All estimations from (\ref{eq:2.2})
coincide with results from general system (\ref{eq:2.1}) within
errors with exception of the $\lambda_{\footnotesize{G}}$ for CMS
at $\sqrt{s_{\footnotesize{NN}}}=7$ TeV. For last case the
estimations from (\ref{eq:2.1}) and (\ref{eq:2.2}) coincide with
each other within 2$\sigma$. Thus the ultimate relations
(\ref{eq:2.2}) provide reasonable estimations for 1D BEC
parameters in both cases of the Cauchy and the Gaussian
distributions within features of modern experiments under
consideration, i.e. at $q_{1} \sim 10^{-2}$ GeV/$c$, $q_{2} \sim
2$ GeV/$c$ and $\alpha_{\footnotesize{L}} \sim 0.8$ which is close
to the region of L\'{e}vy index values with weaker changing of 1D
BEC parameters (Fig.\,\ref{fig:2}).

\textbf{2.~Direction of calculations $\Omega_{\footnotesize{C/G}}
\to \Omega_{\footnotesize{L}}$.}~~ Here the set of 1D BEC
parameters $\{\lambda_{\footnotesize{L}},R_{\footnotesize{L}}\}$
is estimated at \emph{a priori} given $\alpha_{\footnotesize{L}}$
with help of the system of equations (\ref{eq:2.1}) for
experimentally known $\Omega_{\footnotesize{C}}$
\cite{CMS-JHEP-0511-029-2011,ATLAS-Astalos-PhD2015},
$\Omega_{\footnotesize{G}}$
\cite{CMS-PRL-105-032001-2010,ATLAS-Astalos-PhD2015} and finite
$q$-ranges. The estimations for parameters of
$\Omega_{\footnotesize{L}}$ are shown in Table \ref{tab:4.1b}
together with the available experimental results and the values of
$\lambda_{\footnotesize{L}}$, $R_{\footnotesize{L}}$ derived from
experimental analysis with Cauchy distribution are shown on the
first line, the second one corresponds to the calculations with
data for Gaussian distribution for certain experiment at given
energy. The relatively large systematic uncertainties for ATLAS
are driven by corresponding error for L\'{e}vy index
\cite{ATLAS-Astalos-PhD2015}. There is a remarkable agreement
between results of calculations and experimental analyses (Table
\ref{tab:4.1b}): estimations for all 1D BEC parameters coincide
with corresponding experimental values within statistical errors
with exception of the $\lambda_{\footnotesize{L}}$ at
$\sqrt{s_{\footnotesize{NN}}}=7$ TeV for ATLAS data. For last case
the coincidence between estimations from (\ref{eq:2.1}) experiment
is achieved within total errors. This conclusion is for both
$\Omega_{\footnotesize{C}} \to \Omega_{\footnotesize{L}}$ and
$\Omega_{\footnotesize{G}} \to \Omega_{\footnotesize{L}}$ schemes
of calculations. Thus the system of general equations
(\ref{eq:2.1}) provides the high-quality estimations of 1D BEC
parameters for general view centrally-symmetric L\'{e}vy
parametrization $\Omega_{\footnotesize{L}}$ based on the \emph{a
priori} known $\alpha_{\footnotesize{L}}$ and results for
exponential / Gaussian function. One can note the ultimate
relations (\ref{eq:2.2}) for semi-infinite range of Lorentz
invariant quantity $q$ result in reasonable estimations for the
set of 1D BEC parameters
$\{\lambda_{\footnotesize{L}},R_{\footnotesize{L}}\}$ with help of
results for exponential function $\Omega_{\footnotesize{C}}$ as
well as for Gaussian one $\Omega_{\footnotesize{G}}$. Nevertheless
the general system (\ref{eq:2.1}) allows the noticeable
improvement of the results with respect of the (\ref{eq:2.2}) for
chaoticity $\lambda_{\footnotesize{L}}$ derived from results for
Gaussian distribution. This feature can be expected from
Fig.\ref{fig:2}d because curve calculated with (\ref{eq:2.1})
differs from the corresponding ultimate $\alpha_{2}$-dependence at
values $\alpha_{1} \lesssim 0.8$ which are close to the
experimental data (Table \ref{tab:app2a}).

\begin{table*}[h!]
\caption{Parameter values for exponential and Gaussian
parameterizations $\Omega_{\footnotesize{C(G)}}$} \label{tab:4.1}
\begin{center}
\begin{tabular}{cccccccc}
\hline \multicolumn{1}{c}{Collision} &
\multicolumn{1}{c}{$\sqrt{s_{\footnotesize{NN}}}$,} &
\multicolumn{1}{c}{Experiment} & \multicolumn{2}{c}{Estimation
based on the (\ref{eq:2.1})} &
\multicolumn{3}{c}{Experimental values} \rule{0pt}{10pt}\\
\cline{4-8} & GeV & & $\lambda$ & $R$, fm & $\lambda$ &
$R$, fm & Ref. \rule{0pt}{10pt}\\
\hline \rule{0pt}{10pt}
$p+p$ & 900  & CMS & $0.62 \pm 0.08$ & $1.47 \pm 0.24$ & $0.616 \pm 0.011 \pm 0.029$ & $1.56 \pm 0.02 \pm 0.15$ & \cite{CMS-JHEP-0511-029-2011} \\
 &  & & $0.35 \pm 0.06$ & $0.81 \pm 0.22$ & $0.32 \pm 0.01$ & $0.98 \pm 0.03$ & \cite{CMS-PRL-105-032001-2010} \\
& 7000 & ATLAS & $0.73 \pm 0.03 \pm 0.54$ & $2.02 \pm 0.11 \pm 1.79$ & $0.701 \pm 0.006 \pm 0.067$ & $2.021 \pm 0.012 \pm 0.281$& \cite{ATLAS-Astalos-PhD2015} \\
&  & & $0.39 \pm 0.01 \pm 0.20$ & $1.07 \pm 0.04 \pm 0.82$ & $0.302 \pm 0.002 \pm 0.019$ & $1.046 \pm 0.005 \pm 0.114$& \\
&  & CMS & $0.62 \pm 0.06$ & $1.81 \pm 0.23$ & $0.618 \pm 0.009 \pm 0.042$ & $1.89 \pm 0.02 \pm 0.21$ & \cite{CMS-JHEP-0511-029-2011} \\
& & & $0.35 \pm 0.03$ & $0.96 \pm 0.12$ & -- & -- & -- \\
\hline
\end{tabular}
\end{center}
\end{table*}

\begin{table*}[h!]
\caption{Parameter values for general L\'{e}vy parametrization
$\Omega_{\footnotesize{L}}$ at given $\alpha_{\footnotesize{L}}$}
\label{tab:4.1b}
\begin{center}
\begin{tabular}{cccccccc}
\hline \multicolumn{1}{c}{Collision} &
\multicolumn{1}{c}{$\sqrt{s_{\footnotesize{NN}}}$,} &
\multicolumn{1}{c}{Experiment} & \multicolumn{2}{c}{Estimation
based on the (\ref{eq:2.1})} &
\multicolumn{3}{c}{Experimental values} \rule{0pt}{10pt}\\
\cline{4-8} & GeV & & $\lambda$ & $R$, fm & $\lambda$ &
$R$, fm & Ref. \rule{0pt}{10pt}\\
\hline \rule{0pt}{10pt}
$p+p$ & 900  & CMS & $0.85 \pm 0.04 \pm 0.05$ & $2.33 \pm 0.18 \pm 0.23$ & $0.85 \pm 0.06$ & $2.20 \pm 0.17$ & \cite{CMS-JHEP-0511-029-2011} \\
 &  & & $0.89 \pm 0.10$ & $3.2 \pm 0.5$ & $0.93 \pm 0.11$ & $2.5 \pm 0.4$ & \cite{CMS-PRL-105-032001-2010} \\
& 7000 & ATLAS & $0.973 \pm 0.012 \pm 0.332$ & $2.97 \pm 0.06 \pm 1.27$ & $1.02 \pm 0.03 \pm 0.41$ & $2.96 \pm 0.09 \pm 1.31$& \cite{ATLAS-Astalos-PhD2015} \\
&  & & $0.774 \pm 0.010 \pm 0.250$ & $2.86 \pm 0.06 \pm 1.21$ & & & \\
&  & CMS & $0.89 \pm 0.04 \pm 0.07$ & $2.96 \pm 0.17 \pm 0.34$ & $0.90 \pm 0.05$ & $2.83 \pm 0.18$ & \cite{CMS-JHEP-0511-029-2011} \\
\hline
\end{tabular}
\end{center}
\end{table*}

\subsection{Relations between parameters for Cauchy and Gaussian distributions}\label{subsec:4.2}
The system of equations (\ref{eq:3.2}) derived above is used for
estimation of the 1D BEC parameter values for Gaussian
parametrization based on the \emph{a priori} known values for set
$\{\lambda_{\footnotesize{C}},R_{\footnotesize{C}}\}$ of BEC
parameters for exponential parametrization, experimental ranges on
$q$ and vice versa. In the subsection the separation is used on
various multiplicity event classes in $p+p$, $\bar{p}+p$
collisions for 1D BEC results in some experiments. The results for
minimum bias events are shown on the first line, for high
multiplicity events -- on the second line for certain experiment
at given energy in Tables \ref{tab:4.2} and \ref{tab:4.3}.

%\subsubsection{Direction of calculations $\Omega_{\footnotesize{C}} \to \Omega_{\footnotesize{G}}$}\label{subsubsec:4.2.1}
\textbf{1.~Direction of calculations $\Omega_{\footnotesize{C}}
\to \Omega_{\footnotesize{G}}$.}~~Parameters for Gaussian function
are calculated with help of system (\ref{eq:3.2}) and \emph{a
priori} known set
$\{\lambda_{\footnotesize{C}},R_{\footnotesize{C}}\}$. The results
are shown in the Table \ref{tab:4.2} together with available
published experimental results for the Gaussian set
$\{\lambda_{\footnotesize{G}},R_{\footnotesize{G}}\}$. As seen
from the Table \ref{tab:4.2} the estimations for the set of the
Gaussian parameters are equal for published results within (total)
errors for proton-proton collisions with exception of the value of
strength of correlations $\lambda_{\footnotesize{G}}$ in ATLAS
minimum bias events at $\sqrt{s_{\footnotesize{NN}}}=7$ TeV and
CMS result at collision energy $\sqrt{s_{\footnotesize{NN}}}=2.36$
TeV. In the two last cases the agreement is observed within
2$\sigma$. The similar situation is for symmetric nucleus-nucleus
collisions, i.e. the estimations within the present paper for set
of Gaussian parameters
$\{\lambda_{\footnotesize{G}},R_{\footnotesize{G}}\}$ agree with
the results of the WA98 experiment \cite{WA98-EPJC-16-445-2000}
within 2$\sigma$. But there is qualitative agreement only between
results of calculations with help of (\ref{eq:3.3}) and
experimental data for $\bar{p}+p$ collisions
\cite{CPLEAR-ZPC-63-541-1994}. Perhaps, this discrepancy is
dominated by some features of experiment provided unusually large
values of chaoticity for both the exponential and the Gaussian
parameterizations of 1D CF $\mathbf{C}_{2}(q)$. For asymmetric
nuclear interactions the agreement between results of calculations
in the present paper and available experimental data is achieved
mostly within errors for both the $\lambda_{\footnotesize{G}}$ and
the 1D BEC radius. Only estimations for Gaussian 1D radius
$R_{\footnotesize{G}}$ in $\mbox{O}+\mbox{Ag}$ and for
$\lambda_{\footnotesize{G}}$ in $\mbox{O}+\mbox{Au}$ coincide with
corresponding results of the WA80 experiment
\cite{Albrecht-ZPC-53-225-1992} within 2$\sigma$. It should be
emphasized that approximate calculations demonstrate the same
results as in Table \ref{tab:4.2} within errors for all
consecutive simplifications (\ref{eq:3.3}) -- (\ref{eq:3.6}) which
are valid and can be applied for certain experiment. One can note
in particular that as expected the ultimate relations
(\ref{eq:3.1}) work rather well for the CMS results at
$\sqrt{s_{\footnotesize{NN}}}=2.76$ TeV with low enough $q_{1}
\approx 0.6$ MeV/$c$ and high enough $q_{2} \approx 2.0$ GeV/$c$.
Thus detail calculations for case $\Omega_{\footnotesize{C}} \to
\Omega_{\footnotesize{G}}$ confirm both the correctness of
suggestions made above for certain experiments and the validity of
corresponding systems of equations (\ref{eq:3.1}) --
(\ref{eq:3.6}).

%\subsubsection{Direction of calculations $\Omega_{\footnotesize{G}} \to \Omega_{\footnotesize{C}}$}\label{subsubsec:4.2.2}
\textbf{2.~Direction of calculations $\Omega_{\footnotesize{G}}
\to \Omega_{\footnotesize{C}}$.}~~Values of 1D BEC parameters for
exponential function are estimated with help of system
(\ref{eq:3.2}) and \emph{a priori} known values for Gaussian BEC
quantities $\{\lambda_{\footnotesize{G}},R_{\footnotesize{G}}\}$.
The results are presented in Table \ref{tab:4.3} together with
available published experimental results for the set of BEC
parameters $\{\lambda_{\footnotesize{C}},R_{\footnotesize{C}}\}$
corresponded to the Cauchy distribution function in coordinate
space for particle emission points. For $p+p$ collisions there is
agreement between estimations for parameters for exponential
parametrization of the CF $\mathbf{C}_{2}(q)$ calculated with
(\ref{eq:3.2}) and experimental results within (total) errors with
exception of the 1D BEC radius $R_{\footnotesize{C}}$ in CMS at
$\sqrt{s_{\footnotesize{NN}}}=0.9$ TeV. In the last case results
from calculation and experiment coincide within 2$\sigma$. The
similar situation is observed for nucleus-nucleus collisions:
estimations for parameters of exponential function
$\Omega_{\footnotesize{C}}$ derived with (\ref{eq:3.2}) agree with
corresponding experimental results mostly within 1$\sigma$, but
the coincidence is achieved within 2$\sigma$ for
$R_{\footnotesize{C}}$ in both the $\mbox{Pb}+\mbox{Pb}$
collisions at $\sqrt{s_{\footnotesize{NN}}}=17.3$ GeV
\cite{WA98-EPJC-16-445-2000} and the $\mbox{O}+\mbox{Ag}$
reactions at $\sqrt{s_{\footnotesize{NN}}}=19.4$ GeV
\cite{Albrecht-ZPC-53-225-1992}. The estimations of 1D BEC
parameters $\lambda_{\footnotesize{C}}$, $R_{\footnotesize{C}}$
obtained for $\bar{p}+p$ with help of (\ref{eq:3.3}) are in
qualitative agreement with corresponding experimental results
\cite{CPLEAR-ZPC-63-541-1994} even for the case of unusually large
chaoticity. Thus the system of equations (\ref{eq:3.2}) provides
quite reasonable estimations for parameters for exponential
function based on the \emph{a priori} known values of 1D BEC
observables for Gaussian function in various strong interaction
processes at all available experimental energies.

In summary of the section, the systems of equations
(\ref{eq:2.1}), (\ref{eq:3.2}) provide correct estimations for
both desired BEC parameters, namely, the strength of correlations
and the 1D radius in the case of centrally-symmetric L\'{e}vy
distribution $\Omega_{\footnotesize{L}}$ as well as for specific
Cauchy and Gaussian ones. In general the estimations show
remarkable agreement with available experimental data. Thus the
systems of equations suggested in Sec. \ref{sec:2}, \ref{sec:3}
can be useful in experimental data analysis as well as in
phenomenological study for estimation of unknown values of BEC
parameters for some parametrization (\ref{eq:1.1}) of the 1D CF
$\mathbf{C}_{2}(q)$ based on the available values of $\lambda$ and
$R$ for another centrally-symmetric L\'{e}vy distribution. As seen
from Tables \ref{tab:4.1} -- \ref{tab:4.3} the new estimations are
obtained for 1D BEC parameters in $\Omega_{\footnotesize{C}}$,
$\Omega_{\footnotesize{G}}$ in many cases for which the
corresponding experimental results are absent. Thus the systems of
equations derived within the framework of this paper allow the
expansion of the available ensemble of values for 1D BEC
parameters $\lambda$ and $R$ which it is useful for future
investigations.

\begin{table*}[h!]
\caption{Parameter values for Gaussian function
$\Omega_{\footnotesize{G}}$ in (\ref{eq:1.1})} \label{tab:4.2}
\begin{center}
\begin{tabular}{cccccccc}
\hline \multicolumn{1}{c}{Collision} &
\multicolumn{1}{c}{$\sqrt{s_{\footnotesize{NN}}}$,} &
\multicolumn{1}{c}{Experiment} & \multicolumn{2}{c}{Estimation
based on the (\ref{eq:3.2})} &
\multicolumn{3}{c}{Experimental values} \rule{0pt}{10pt}\\
\cline{4-8} & GeV & & $\lambda_{\footnotesize{G}}$ &
$R_{\footnotesize{G}}$, fm & $\lambda_{\footnotesize{G}}$ &
$R_{\footnotesize{G}}$, fm & Ref. \rule{0pt}{10pt}\\
\hline \rule{0pt}{10pt}
$p+p$ & 63.0 & AFS & $0.47 \pm 0.04$ & $0.73 \pm 0.07$ & $0.40 \pm 0.03$ & $0.82 \pm 0.05$ & \cite{Akesson-ZPC-36-517-1987} \\
& 900  & ALICE & $0.305 \pm 0.024$ & $1.00 \pm 0.09^{+0.06}_{-0.18}$ & $0.35 \pm 0.03$ & $1.00 \pm 0.06^{+0.10}_{-0.20}$ & \cite{ALICE-PRD-82-052001-2010} \\
&   &  & $0.357 \pm 0.025$ & $0.89 \pm 0.06^{+0.08}_{-0.18}$ & $0.310 \pm 0.026$ & $1.18 \pm 0.09^{+0.07}_{-0.17}$ & \\
&   & ATLAS & $0.41 \pm 0.01 \pm 0.03$ & $1.00 \pm 0.03 \pm 0.08$& $0.34 \pm 0.01 \pm 0.03$ & $1.00 \pm 0.03 \pm 0.08$ & \cite{Sykora-DIS-2015} \\
&   & CMS & $0.351 \pm 0.006 \pm 0.013$ & $0.83 \pm 0.01 \pm 0.08$ & $0.32 \pm 0.01$ & $0.98 \pm 0.03$ & \cite{CMS-PRL-105-032001-2010} \\
& 2360 & & $0.348 \pm 0.030 \pm 0.013$ & $1.04 \pm 0.08 \pm 0.10$ & $0.32 \pm 0.01$ & $0.98 \pm 0.03$ & \\
% !!! ATTENTION !!!
% The values of \lambda_{C} and R_{C} INTEGRATED over k_{T} and N_{ch} are used as input parameter for 2.76 TeV
& 2760 & & $0.366 \pm 0.005 \pm 0.025$ & $0.915 \pm 0.007 \pm 0.116$ & -- & -- & -- \\
& 7000 & ALICE & $0.719 \pm 0.002 \pm 0.047$ & $1.148 \pm 0.007^{+0.04}_{-0.02} $ & $0.645 \pm 0.003 \pm 0.047$ & $1.430 \pm 0.005^{+0.16}_{-0.30}$& \cite{ALICE-PLB-739-139-2014} \\
&  & ATLAS & $0.381 \pm 0.003 \pm 0.022$ & $1.092 \pm 0.005 \pm 0.074$ & $0.327 \pm 0.002 \pm 0.020$ & $1.130 \pm 0.005 \pm 0.086$& \cite{Sykora-DIS-2015} \\
&  &  & $0.266 \pm 0.009 \pm 0.015$ & $1.25 \pm 0.03 \pm 0.09$ & $0.251 \pm 0.010 \pm 0.018$ & $1.38 \pm 0.04 \pm 0.12$& \\
&  & CMS & $0.344 \pm 0.005 \pm 0.018$ & $1.00 \pm 0.01 \pm 0.10$ & -- & -- & -- \\
\hline \rule{0pt}{10pt}
$\bar{p}+p$& 1.89 & CPLEAR & $2.332 \pm 0.025$  & $0.972 \pm 0.014$ & $1.96 \pm 0.03$ & $1.04 \pm 0.01$ & \cite{CPLEAR-ZPC-63-541-1994} \rule{0pt}{12pt}\\
\hline \rule{0pt}{10pt}
$p+\mbox{Pb}$& 5020 & CMS & $0.358 \pm 0.007 \pm 0.021$ & $1.70 \pm 0.02 \pm 0.12$ & -- & -- & -- \rule{0pt}{12pt}\\
\hline \rule{0pt}{10pt}
$\mbox{Pb}+\mbox{Pb}$& 17.3 & WA98 & $0.327 \pm 0.008$ & $6.51 \pm 0.10$ & $0.307 \pm 0.008$ & $6.83 \pm 0.10$ & \cite{WA98-EPJC-16-445-2000} \rule{0pt}{12pt}\\
\hline \rule{0pt}{10pt}
\hspace*{-0.1cm}$\mbox{O}+\mbox{C}$& 19.4 & WA80 & $0.44 \pm 0.05$ & $2.8 \pm 0.3$ & $0.40 \pm 0.03$ & $2.90 \pm 0.21$ & \cite{Albrecht-ZPC-53-225-1992} \\
$\mbox{O}+\mbox{Cu}$& & & $0.24 \pm 0.07$ & $2.53 \pm 0.11$ & $0.17 \pm 0.03$ & $2.35 \pm 0.11$ &  \\
$\mbox{O}+\mbox{Ag}$& & & $0.28 \pm 0.10$ & $2.71 \pm 0.11$ & $0.17 \pm 0.04$ & $2.44 \pm 0.11$ &  \\
$\mbox{O}+\mbox{Au}$& & & $0.110 \pm 0.015$ & $1.63 \pm 0.05$ & $0.085 \pm 0.007$ & $1.68 \pm 0.06$ &  \\
\hline
\end{tabular}
\end{center}
\end{table*}

\begin{table*}[h!]
\caption{Parameter values for exponential function
$\Omega_{\footnotesize{C}}$ in (\ref{eq:1.1})} \label{tab:4.3}
\begin{center}
\begin{tabular}{cccccccc}
\hline \multicolumn{1}{c}{Collision} &
\multicolumn{1}{c}{$\sqrt{s_{\footnotesize{NN}}}$,} &
\multicolumn{1}{c}{Experiment} & \multicolumn{2}{c}{Estimation
based on the (\ref{eq:3.2})} &
\multicolumn{3}{c}{Experimental values} \rule{0pt}{10pt}\\
\cline{4-8} & GeV & & $\lambda_{\footnotesize{C}}$ &
$R_{\footnotesize{C}}$, fm & $\lambda_{\footnotesize{C}}$ &
$R_{\footnotesize{C}}$, fm & Ref. \rule{0pt}{10pt}\\
\hline \rule{0pt}{10pt}
$p+p$ & 7.21 & E766  & $1.63 \pm 0.10$   & $2.29 \pm 0.08$ & -- & -- & -- \\
& 26.0      & NA23  & $1.4 \pm 0.7$     & $2.6 \pm 0.7$   & -- & -- & -- \\
& 63.0      & AFS   & $0.67 \pm 0.05$   & $1.50 \pm 0.10$ & $0.77 \pm 0.07$ & $1.32 \pm 0.13$ & \cite{Akesson-ZPC-36-517-1987} \\
& 200       & STAR  & $0.588 \pm 0.010$ & $2.43 \pm 0.04 \pm 0.26$   & -- & -- & -- \\
& 900       & ALICE & $0.63 \pm 0.06$   & $1.89 \pm 0.12^{+0.20}_{-0.40}$ & $0.63 \pm 0.05$ & $1.67 \pm 0.12^{+0.16}_{-0.35}$ & \cite{ALICE-PRD-82-052001-2010} \\
&           &       & $0.57 \pm 0.06$   & $2.26 \pm 0.19^{+0.14}_{-0.34}$ & $0.55 \pm 0.05$ & $1.90 \pm 0.18^{+0.11}_{-0.36}$ & \\
&           & ATLAS & $0.62 \pm 0.03 \pm 0.08$ & $1.84 \pm 0.07 \pm 0.20$& $0.74 \pm 0.03 \pm 0.09$ & $1.83 \pm 0.07 \pm 0.20$ & \cite{Sykora-DIS-2015} \\
&           & CMS   & $0.57 \pm 0.02$   & $1.85 \pm 0.06$ & $0.616 \pm 0.011 \pm 0.029$ & $1.56 \pm 0.02 \pm 0.15$ & \cite{CMS-PRL-105-032001-2010,CMS-JHEP-0511-029-2011} \\
& 2360      &       & $0.60 \pm 0.03$   & $1.86 \pm 0.07$ & $0.66 \pm 0.07 \pm 0.05$ & $1.99 \pm 0.18 \pm 0.24$ & \cite{CMS-PRL-105-032001-2010} \\
& 7000      & ALICE & $1.104 \pm 0.006 \pm 0.112$ & $2.627 \pm 0.010^{+0.320}_{-0.626}$ & $1.180 \pm 0.005 \pm 0.084$ & $2.038 \pm 0.014^{+0.083}_{-0.046}$& \cite{ALICE-PLB-739-139-2014} \\
&           & ATLAS & $0.627 \pm 0.005 \pm 0.058$ & $2.163 \pm 0.012 \pm 0.209$ & $0.718 \pm 0.006 \pm 0.062$ & $2.067 \pm 0.012 \pm 0.182$& \cite{Sykora-DIS-2015} \\
&           &       & $0.521 \pm 0.027 \pm 0.055$ & $2.76 \pm 0.10 \pm 0.29$ & $0.531 \pm 0.024 \pm 0.046$ & $2.46 \pm 0.08 \pm 0.22$& \\
\hline \rule{0pt}{10pt}
$\bar{p}+p$& 1.89 & CPLEAR & $4.21 \pm 0.09$ & $2.079 \pm 0.028$ & $4.79 \pm 0.10$ & $1.89 \pm 0.04$ & \cite{CPLEAR-ZPC-63-541-1994} \rule{0pt}{12pt}\\
\hline \rule{0pt}{10pt}
$\mbox{Au}+\mbox{Au}$& 4.86 & E802 & $0.86 \pm 0.08$ & $12.4 \pm 0.6$ & -- & -- & -- \rule{0pt}{12pt}\\
$\mbox{Pb}+\mbox{Pb}$& 17.3 & NA44 & $1.06 \pm 0.10$ & $15.1 \pm 0.9$ & -- & -- & -- \\
                     &      & WA98 & $0.69 \pm 0.02$ & $14.1 \pm 0.3$ & $0.718 \pm 0.023$ & $13.34 \pm 0.26$ & \cite{WA98-EPJC-16-445-2000}\\
$\mbox{Au}+\mbox{Au}$& 130   & STAR & $0.99 \pm 0.03 \pm 0.08$ & $13.0 \pm 0.3 \pm 0.9$ & -- & -- & -- \\
\hline \rule{0pt}{10pt}
\hspace*{-0.2cm}$\mbox{Si}+\mbox{Al}$& 5.41   & E802 & $1.23 \pm 0.08$ & $8.4 \pm 0.3$ & -- & -- & -- \\
$\mbox{Si}+\mbox{Au}$& & & $0.95 \pm 0.05$ & $9.4 \pm 0.3$ & -- & -- & -- \\
$\mbox{S}+\mbox{Pb}$& 17.3 & NA44 & $0.81 \pm 0.05$ & $7.8 \pm 0.6$ & -- & -- & -- \\
\hspace*{-0.1cm}$\mbox{O}+\mbox{C}$& 19.4 & WA80 & $0.85 \pm 0.08$ & $5.9 \pm 0.5$ & $0.92 \pm 0.13$ & $5.7 \pm 0.7$ & \cite{Albrecht-ZPC-53-225-1992} \\
$\mbox{O}+\mbox{Cu}$& & & $0.34 \pm 0.06$ & $4.7 \pm 0.2$ & $0.49 \pm 0.14$ & $5.05 \pm 0.25$ & \\
$\mbox{O}+\mbox{Ag}$& & & $0.34 \pm 0.09$ & $4.9 \pm 0.2$ & $0.59 \pm 0.21$ & $5.46 \pm 0.24$ & \\
$\mbox{O}+\mbox{Au}$& & & $0.156 \pm 0.014$ & $3.19 \pm 0.13$ & $0.20 \pm 0.03$ & $3.07 \pm 0.12$ & \\
\hline
\end{tabular}
\end{center}
\end{table*}

%#5
\section{Summary}\label{sec:5}

The case is investigated for smooth approximation of the one
experimental 1D Bose\,--\,Einstein correlation function by two
various centrally-symmetric L\'{e}vy parameterizations. It is
suggested that lowest moments of corresponding distributions are
equal approximately. Then the relations are derived between sets
of 1D BEC observables, namely, strength of correlations and source
radius, for two general view centrally-symmetric L\'{e}vy
parameterizations under consideration for the first time. The
relations obtained in the paper take into account the finiteness
of range of Lorentz invariant four-momentum difference in
experimental studies. Detailed analysis results in the systems of
transcendental equations for various finite ranges of the Lorentz
invariant four-momentum difference in the specific case of the
exponential and Gaussian parameterizations for correlation
function. It is shown that finite range of $q$ should be taken
into account and corresponding systems of equations should be used
for derivation of set $\{\lambda, R\}$ based on the \emph{a
priori} known values of corresponding parameters for both cases
the two general view centrally-symmetric L\'{e}vy
parameterizations and the two specific functions (exponential and
Gaussian) most used in experimental studies. The ultimate
relations derived for semi-infinite range of $q$ can be utilized
carefully for experimental analysis and these equations can
produce the reasonable estimations for 1D BEC parameters for
ranges of L\'{e}vy indexes $\forall\,i=1,2: \alpha_{i} \gtrsim 1$
only. Furthermore it is demonstrated that the corresponding
ultimate relations for specific case of Cauchy and Gaussian
distributions for source in coordinate space produce the
reasonable estimations of 1D BEC parameters for few modern
experimental analyses only. The mathematical formalism suggested
within the framework of the preset paper is verified with help of
experimental results obtained for wide set of strong interaction
processes in all available energy range. The two pairs of
distributions are considered: general view centrally-symmetric
L\'{e}vy one with specific case (Cauchy / Gaussian); two specific
Cauchy and Gaussian distributions. For both cases verifications
are made for both directions of calculations. Namely, the
calculations have been made for estimation of 1D BEC observables
for Cauchy / Gaussian function based on the \emph{a priori} known
values of parameters of general L\'{e}vy parametrization and vice
versa; for estimations of Gaussian parameters based on the \emph{a
priori} known values of observables for Cauchy distribution and
vice versa. Comparison shows the quantitative agreement between
estimations derived with help of mathematical formalism developed
in the paper and most of available experimental results for both
pairs consisting of the general view centrally-symmetric L\'{e}vy
parametrization and specific (exponential / Gaussian) function and
the two specific source distributions (Cauchy and Gaussian) most
used in experimental studies for any direction of calculations.

\section*{Acknowledgments}

The author is grateful to Prof. G. Alexander for useful
discussions.

%#6
\appendix
%Appendix A
\section{Data for 1D BEC parameters in strong interactions}\label{sec:6}

\hspace*{0.5cm}In this Appendix experimental database is shown in
Tables \ref{tab:app1a}, \ref{tab:app2a} for 1D BEC parameters for
identical charged pions produced in $p+p$, $\bar{p}+p$ and
$\mbox{A}_{1}+\mbox{A}_{2}$ interactions\footnote{In Table
\ref{tab:app1a} total uncertainties are shown for exponential
parametrization in CPLEAR \cite{CPLEAR-ZPC-63-541-1994} and for
NA44 experiment \cite{NA44-PRC-58-1656-1998}.}. Some of the
numerical values used in the Sec.~\ref{sec:4}. The pion pairs with
low average transverse momentum, $\langle k_{T}\rangle$, are
considered for all types of strong interaction processes. The
results for most central nucleus-nucleus collisions are used and
as consequence additional separation is made for $p+p$ collisions
on minimum bias and high multiplicity events if corresponding
experimental 1D BEC values are available. In the last case the
results for minimum bias events are shown on the first line, for
high multiplicity events -- on the second line for certain
experiment at given energy.

\begin{table*}[h!]
\caption{Experimental results for special cases of parametrization
(\ref{eq:1.1})} \label{tab:app1a}
\begin{center}
\begin{tabular}{cccccccc}
\hline \multicolumn{1}{c}{Collision} &
\multicolumn{1}{c}{$\sqrt{s_{\footnotesize{NN}}}$,} &
\multicolumn{1}{c}{Experiment} & \multicolumn{2}{c}{Exponential
function $\Omega_{\footnotesize{C}}$} &
\multicolumn{2}{c}{Gaussian function $\Omega_{\footnotesize{G}}$}
&
\multicolumn{1}{c}{Ref.} \rule{0pt}{10pt}\\
\cline{4-8} & GeV & & $\lambda_{\footnotesize{C}}$ &
$R_{\footnotesize{C}}$, fm & $\lambda_{\footnotesize{G}}$ &
$R_{\footnotesize{G}}$, fm & \rule{0pt}{10pt}\\
\hline \rule{0pt}{10pt}
$p+p$ & 7.21 & E766 & -- & -- & $0.466 \pm 0.015$ & $0.95 \pm 0.03$ & \cite{Uribe-PRD-49-4373-1994} \\
& 26.0      & NA23  & -- & -- & $0.32 \pm 0.08$ & $1.02 \pm 0.20$ & \cite{NA23-ZPC-43-341-1989} \\
& 27.4      & NA27  & -- & -- & -- & $1.20 \pm 0.03$ & \cite{NA27-ZPC-54-21-1992} \\
& 31.0      & ABCDHW  & -- & -- & $0.35 \pm 0.04$ & $1.01 \pm 0.08$ & \cite{Breakstone-ZPC-33-333-1987} \\
& 44.0      &   & -- & -- & $0.42 \pm 0.04$ & $1.13 \pm 0.07$ & \\
& 62.0      &   & -- & -- & $0.42 \pm 0.08$ & $1.69 \pm 0.25$ & \\
& 63.0      & AFS   & $0.77 \pm 0.07$   & $1.32 \pm 0.13$ & $0.40 \pm 0.03$ & $0.82 \pm 0.05$ & \cite{Akesson-ZPC-36-517-1987} \\
& 200       & STAR  & --   & -- & $0.345 \pm 0.005$ & $1.32 \pm 0.02 \pm 0.13$ & \cite{STAR-PRC-83-064905-2011} \\
& 900       & ALICE & $0.63 \pm 0.05$   & $1.87 \pm 0.12^{+0.16}_{-0.35}$ & $0.35 \pm 0.03$   & $1.00 \pm 0.06^{+0.10}_{-0.20}$ & \cite{ALICE-PRD-82-052001-2010} \\
&           &       & $0.55 \pm 0.05$   & $1.90 \pm 0.18^{+0.11}_{-0.36}$ & $0.310 \pm 0.026$ & $1.184 \pm 0.092^{+0.067}_{-0.168}$ & \\
&           & ATLAS & $0.74 \pm 0.03 \pm 0.09$ & $1.83 \pm 0.07 \pm 0.20$& $0.34 \pm 0.01 \pm 0.03$ & $1.00 \pm 0.03 \pm 0.08$ & \cite{Sykora-DIS-2015} \\
&           & CMS   & $0.616 \pm 0.011 \pm 0.029$ & $1.56 \pm 0.02 \pm 0.15$ & $0.32 \pm 0.01$ & $0.98 \pm 0.03$ & \cite{CMS-PRL-105-032001-2010,CMS-JHEP-0511-029-2011} \\
& 2360      &       & $0.66 \pm 0.07 \pm 0.05$    & $1.99 \pm 0.18 \pm 0.24$ & $0.32 \pm 0.01$ & $0.98 \pm 0.03$ & \cite{CMS-PRL-105-032001-2010} \\
& 2760      &       & $0.808 \pm 0.017 \pm 0.062$ & $2.35 \pm 0.07 \pm 0.31$ & -- & -- & \cite{Padula-WPCF-2014} \\
& 7000      & ALICE & $1.180 \pm 0.005 \pm 0.084$ & $2.038 \pm 0.014^{+0.083}_{-0.046}$ & $0.645 \pm 0.003 \pm 0.047$ & $1.430 \pm 0.005^{+0.158}_{-0.300}$& \cite{ALICE-PLB-739-139-2014} \\
&           & ATLAS & $0.718 \pm 0.006 \pm 0.062$ & $2.067 \pm 0.012 \pm 0.182$ & $0.327 \pm 0.002 \pm 0.020$ & $1.130 \pm 0.005 \pm 0.086$& \cite{Sykora-DIS-2015} \\
&           &       & $0.531 \pm 0.024 \pm 0.046$ & $2.46 \pm 0.08 \pm 0.22$ & $0.251 \pm 0.010 \pm 0.018$ & $1.38 \pm 0.04 \pm 0.12$& \\
&           & CMS   & $0.618 \pm 0.009 \pm 0.042$ & $1.89 \pm 0.02 \pm 0.21$ & -- & -- & \cite{CMS-JHEP-0511-029-2011} \\
\hline \rule{0pt}{10pt}
$\bar{p}+p$& 1.89 & CPLEAR & $4.79 \pm 10$ & $1.89 \pm 0.04$ & $1.96 \pm 0.03$ & $1.04 \pm 0.01$ & \cite{CPLEAR-ZPC-63-541-1994} \rule{0pt}{12pt}\\
           & 1800 & E735   & -- & -- & $0.24 \pm 0.02$ & $1.46 \pm 0.10 \pm 0.23$ & \cite{Alexopoulos-PRD-48-1931-1993} \\
           & 1960 & CDF    & $0.89 \pm 0.03$ & $1.67 \pm 0.05$ & $0.50 \pm 0.04$ & $1.79 \pm 0.08$ & \cite{Lovas-PhD-2008} \\
\hline \rule{0pt}{10pt}
$p+\mbox{Pb}$& 5020 & ALICE & $1.230 \pm 0.016^{+0.088}_{-0.141}$ & $4.82 \pm 0.05^{+0.25}_{-0.72}$ & $0.603 \pm 0.006 \pm 0.056$ & $2.780 \pm 0.018^{+0.418}_{-0.668}$ & \cite{ALICE-PLB-739-139-2014} \rule{0pt}{12pt}\\
           &  & ATLAS       & -- & $5.32 \pm 0.06^{+0.50}_{-0.14}$& -- & -- & \cite{ATLAS-note-2015-054-2015} \\
           &  & CMS         & $0.81 \pm 0.02 \pm 0.07$ & $3.55 \pm 0.05 \pm 0.30$ & -- & -- & \cite{Sikler-arXiv-1411.6609-2014} \\
\hline \rule{0pt}{10pt}
$\mbox{Au}+\mbox{Au}$& 4.86 & E802 & -- & -- & $0.44 \pm 0.03$ & $6.32 \pm 0.29$ & \cite{E802-PRC-66-054906-2002} \rule{0pt}{12pt}\\
$\mbox{Au}+\mbox{Pb}$& 17.3 & NA49 & -- & -- & $0.560 \pm 0.023$ & -- & \cite{NA49-EPJC-2-661-1998} \\
$\mbox{Pb}+\mbox{Pb}$&      & NA44 & -- & -- & $0.52 \pm 0.04$ & $7.6 \pm 0.4$ & \cite{NA44-PRC-58-1656-1998}\\
&       & WA98 & $0.718 \pm 0.023$ & $13.34 \pm 0.26$ & $0.307 \pm 0.008$ & $6.83 \pm 0.10$ & \cite{WA98-EPJC-16-445-2000}\\
$\mbox{Au}+\mbox{Au}$& 130   & PHENIX & -- & -- & -- & $6.0 \pm 0.3$ & \cite{PHENIX-PRL-88-192302-2002} \\
                     &       & STAR   & -- & -- & $0.450 \pm 0.009 \pm 0.027$ & $6.30 \pm 0.12 \pm 0.38$ & \cite{STAR-PRL-87-082301-2001} \\
$\mbox{Pb}+\mbox{Pb}$& 2760   & ALICE & $1.830 \pm 0.003 \pm 0.156$ & $19.85 \pm 0.02 \pm 0.28$ & $0.689 \pm 0.001 \pm 0.096$ & $9.70 \pm 0.06 \pm 1.17$ & \cite{ALICE-PLB-739-139-2014} \\

\hline \rule{0pt}{10pt}
\hspace*{-0.2cm}$\mbox{Si}+\mbox{Al}$& 5.41   & E802 & -- & -- & $0.68 \pm 0.04$ & $4.42 \pm 0.16$ & \cite{E802-PRC-66-054906-2002} \\
$\mbox{Si}+\mbox{Au}$& & & -- & -- & $0.511 \pm 0.026$ & $4.91 \pm 0.15$ & \\
$\mbox{S}+\mbox{Pb}$& 17.3 & NA44 & -- & -- & $0.42 \pm 0.02$ & $4.00 \pm 0.27$ & \cite{NA44-PRC-58-1656-1998} \\
\hspace*{-0.1cm}$\mbox{O}+\mbox{C}$& 19.4 & WA80 & $0.92 \pm 0.13$ & $5.7 \pm 0.7$ & $0.40 \pm 0.03$ & $2.90 \pm 0.21$ & \cite{Albrecht-ZPC-53-225-1992} \\
$\mbox{O}+\mbox{Cu}$&  &  & $0.49 \pm 0.14$ & $5.05 \pm 0.25$ & $0.17 \pm 0.03$ & $2.35 \pm 0.11$ & \\
$\mbox{O}+\mbox{Ag}$&  &  & $0.59 \pm 0.21$ & $5.46 \pm 0.24$ & $0.17 \pm 0.04$ & $2.44 \pm 0.11$ & \\
$\mbox{O}+\mbox{Au}$&  &  & $0.20 \pm 0.03$ & $3.07 \pm 0.12$ & $0.085 \pm 0.007$ & $1.68 \pm 0.06$ & \\
                    &  & NA35 & -- & -- & $0.29 \pm 0.03$ & $4.00 \pm 0.20$ & \cite{NA35-PLB-203-320-1988}\\
\hline
\end{tabular}
\end{center}
\end{table*}

\begin{table*}[h!]
\caption{Experimental results for general centrally-symmetric
L\'{e}vy parametrization (\ref{eq:1.1})} \label{tab:app2a}
\begin{center}
\begin{tabular}{ccccccc}
\hline \multicolumn{1}{c}{Collision} &
\multicolumn{1}{c}{$\sqrt{s_{\footnotesize{NN}}}$, GeV} &
\multicolumn{1}{c}{Experiment} & \multicolumn{3}{c}{General view
function $\Omega_{\footnotesize{L}}$} &
\multicolumn{1}{c}{Ref.} \rule{0pt}{10pt}\\
\cline{4-7} & & & $\lambda_{\footnotesize{L}}$ &
$R_{\footnotesize{L}}$, fm & $\alpha_{\footnotesize{L}}$ & \rule{0pt}{10pt}\\
\hline \rule{0pt}{10pt}
$p+p$ & 900 & CMS   & $0.93 \pm 0.11$ & $2.5 \pm 0.4$ & $0.76 \pm 0.06$ & \cite{CMS-PRL-105-032001-2010} \\
& & & $0.85 \pm 0.06$ & $2.20 \pm 0.17$ & $0.81 \pm 0.03$ & \cite{CMS-JHEP-0511-029-2011} \\
& 7000      & ATLAS & $1.02 \pm 0.03 \pm 0.41$ & $2.96 \pm 0.09 \pm 1.31$ & $0.81 \pm 0.01 \pm 0.18$ & \cite{ATLAS-Astalos-PhD2015} \\
&     & CMS   & $0.90 \pm 0.05$ & $2.83 \pm 0.18$ & $0.792 \pm 0.024$ & \cite{CMS-JHEP-0511-029-2011}\\
\hline
\end{tabular}
\end{center}
\end{table*}

\newpage
% Figure 1.
\newpage
% Figure 2.
\newpage
% Figure 3.
\newpage
% Figure 4.
\newpage
% Figure 5.
\newpage
% Figure 6.

\end{document}